%% file: main.tex
\documentclass[conference]{IEEEtran}
\usepackage{cite}
\usepackage{amsmath,amssymb,amsfonts}
\usepackage{algorithmic}
\usepackage{graphicx}
\usepackage{textcomp}
\usepackage{xcolor}
\usepackage{fancyhdr}
\usepackage[hyphens]{url}
\usepackage[caption=false,font=footnotesize]{subfig}
\usepackage[ruled,vlined]{algorithm2e}
\usepackage{tabularx}
\usepackage{threeparttable}
\usepackage{xfrac}
\usepackage{pifont}

\def\BibTeX{{\rm B\kern-.05em{\sc i\kern-.025em b}\kern-.08em
    T\kern-.1667em\lower.7ex\hbox{E}\kern-.125emX}}

\title{Accelerating Markov Random Field Inference with Uncertainty Quantification} 
\author{\IEEEauthorblockN{Ramin Bashizade, Xiangyu Zhang, Sayan Mukherjee, Alvin R. Lebeck}
\IEEEauthorblockA{
\textit{Duke University}\\
Durham, NC, USA \\
ramin@cs.duke.edu, xiangyu.zhang@duke.edu, sayan@stat.duke.edu, alvy@cs.duke.edu}
}

\begin{document}

\maketitle
\pagestyle{plain}

\begin{abstract}

Statistical machine learning methods have widespread applications in various domains.
These methods include probabilistic algorithms, such as Markov Chain Monte-Carlo
(MCMC), which rely on generating random numbers from probability distributions.
These algorithms are computationally expensive on conventional processors, yet their
statistical properties, namely interpretability and uncertainty quantification
compared to deep learning, make them an attractive alternative approach.
Therefore, hardware specialization can be adopted to address the shortcomings
of conventional processors in running these applications.

In this paper, we propose a high-throughput accelerator for first-order Markov Random Field
(MRF) inference, a powerful model for representing a wide range of applications,
using MCMC with Gibbs sampling. We propose a tiled architecture which takes
advantage of near-memory computing, and memory banking and communication
schemes tailored to the semantics of first-order MRF. Additionally, we propose a novel hybrid
on-chip/off-chip memory system and logging scheme to efficiently support uncertainty quantification.
This memory system design is not specific to MRF models and is applicable to applications
using probabilistic algorithms. In addition, it dramatically reduces off-chip memory bandwidth requirements.

We implemented an FPGA prototype of our proposed architecture using
high-level synthesis tools and achieved 146MHz frequency for an accelerator with
32 function units on an Intel Arria 10 FPGA. Compared to prior work on FPGA,
our accelerator achieves 26$\times$ speedup. Furthermore, our proposed memory system and logging scheme to support uncertainty quantification reduces off-chip bandwidth consumption by 71\% for two benchmark applications. ASIC analysis in 15nm technology
node shows our design with 2048 function units running at 3GHz outperforms GPU implementations
of motion estimation and stereo vision run on Nvidia RTX 2080 Ti
by 120$\times$-210$\times$ while occupying only 7.7\% of the area.

\end{abstract}

\maketitle

\input{1-intro.tex}

\input{2-motivation.tex}

\input{3-spa.tex}

\input{4-methodology.tex}

\input{5-evaluation.tex}

\input{6-relatedwork.tex}

\input{7-conclusion.tex}

\section*{Acknowledgements}
This project is supported in part by Intel, the Semiconductor
Research Corporation and the National Science Foundation (CNS-1616947).

\bibliographystyle{IEEEtranS}

\end{document}

%% file: 1-intro.tex
\section{Introduction}
\label{Sec:Introduction}

Statistical machine learning has widespread applications
such as image analysis \cite{MRFVision}, natural language processing \cite{NLP},
global health \cite{Epidemiology}, wireless communications \cite{MIMO}, autonomous driving \cite{AutonomousDriving}, etc. \cite{ComparativeStudyMRF, BlindSignalSeparation, MachineLearningProbabilistic, ProbabilityComputing}.
Many such approaches use probabilistic algorithms,
e.g., Markov chain Monte-Carlo (MCMC), which can be adopted to create
generalized frameworks for solving a wide range of problems.
Compared to Deep Neural Networks, these methods make it easier to gain insight
into why a result is obtained, and to what degree we can be certain about
the results. This can be achieved by quantifying the uncertainty of the result.
This is a very important feature in some contexts, such as many
image segmentation applications where quantifying the uncertainty
in the segmentation boundaries is crucial (e.g., a surgeon's
decision to resect what sections of a tumor will be impacted by the
segmentation generated by an algorithm \cite{LungCT, BrainSegmentation}).

However, the benefits of MCMC come at a price. Since MCMC
requires iteratively sampling from probability
distributions, it is often computationally intensive.
This is due to the significant overhead of
sampling in conventional processors \cite{RSUG}. Furthermore,
MCMC at first appears to be a sequential algorithm because updating each random
variable (RV) depends on the latest value of all other RVs, which
means that it may take a long time to finish. Deploying
pseudo-random number generation can help reduce the
sampling inefficiency \cite{SPU}. To avoid the overhead of serial execution,
though, one can take advantage of the conditional independence of RVs,
i.e., develop a schedule which allows multiple independent
RVs be updated in parallel \cite{EricJonas, ParallelGibbs, GPUAcceleratedGibbs, Gibbs}.

In this paper, we propose an accelerator which builds on these ideas
and fuses them with architectural contributions that allow fast and
efficient execution of MCMC and minimize the overhead of uncertainty
quantification. To be more specific, we propose a tiled architecture
to exploit near-memory computing, and the parallelism exposed by taking
advantage of the conditional independence of RVs in the first-order
Markov Random Field (MRF) model. MRF is a powerful model that can be
utilized to represent a wide range of problems \cite{MRFVision, ComparativeStudyMRF, Stereophonic}.
Our proposed accelerator supports first-order MRF inference using MCMC in two modes of pure sampling or optimization. The main difference is that in optimization mode, the algorithm converges faster whereas in pure sampling mode the exact distribution of the solution is obtained which can be used in uncertainty quantification. Therefore, in this paper we focus on the pure sampling mode.
We develop memory
banking and on-chip communications schemes tailored to the semantics
of the MRF model to facilitate a stall-free pipeline (\textbf{Contribution 1}). Our proposed accelerator targets first-order MRF, and although not general enough for all MCMC applications, it supports an important class of algorithms (stereo vision, image segmentation, motion estimation/optical flow) that are utilized by numerous higher-level computer vision applications. Many of our
proposed techniques are also applicable to other structured probabilistic models. This work takes the first steps toward using these techniques for designing more generalized probabilistic accelerators.

As part of our tiled accelerator architecture, we propose
a novel hybrid on-chip/off-chip memory system to efficiently support uncertainty quantification (\textbf{Contribution 2}).
Probabilistic algorithms solve problems by gradually converging toward the final solution, and thus,
regardless of model or domain, most RVs change labels less often. To quantify this effect,
we carefully analyze the behavior of two image analysis applications \cite{MotionEstimation,StereoVision},
and observe that most RVs take on a limited number of unique labels during
the execution. We build on this insight to design the novel memory system that
stores a few more frequently selected values on-chip, and sends the rest to off-chip
memory to form a log, which is processed at the end of the execution to generate
the histogram of RVs used for uncertainty quantification. This approach strikes
a balance between on-chip memory capacity and off-chip communication
bandwidth. Our experiments on an FPGA prototype for two image analysis applications show that our proposed approach for uncertainty quantification reduces off-chip memory bandwidth usage by 71\% compared to na\"ively storing all RV values at every iteration. Even though we use this solution in the context of first-order MRF
inference, we believe the insight should hold for other types of probabilistic
algorithms accelerators.\footnote{This solution is applicable to
problems that have discrete random variables.}

We implement an FPGA prototype of our proposed design using Intel
High-Level Synthesis (HLS) compiler \cite{HLS}, and developed the necessary
runtime to verify and evaluate our implementation using real-world
applications and input datasets. The results show that our design
achieves a clock rate of 146MHz and a throughput of 4.672B
labels/sec on an Arria 10 FPGA. This is 26$\times$ speedup over
previous work \cite{FPGAGibbs}. We also perform ASIC
analysis on our HLS implementation using Mentor Graphics HLS Compiler \cite{Catapult}
and show that an accelerator with 2048 function units running at 3GHz in 15nm
technology node \cite{Nangate-15} outperforms GPU implementations of motion estimation
and stereo vision on an RTX 2080 Ti by 120$\times$-210$\times$ while only occupying 7.7\% of the area.
Moreover, the aforementioned ASIC design point supports real-time processing of Full-HD images with
64 labels per pixel at 30fps for 1500 iterations per frame. (\textbf{Contribution 3}).

The rest of the paper is organized as follows. Section \ref{Sec:BackgroundAndMotivation} provides
a brief background about probabilistic algorithms and
the motivation for our work. The overview and challenges of
designing the accelerator are discussed in
Section \ref{Sec:DesignOverviewChallenges}. Section \ref{Sec:SPA} describes the
proposed architecture. Sections \ref{Sec:Methodology} and \ref{Sec:Evaluation}
present implementation and evaluation methodology, and
discussion of the results. Related work is reviewed in
Section \ref{Sec:RelatedWork}. Finally, Section \ref{Sec:Conclusion} concludes the paper.

%% file: 2-motivation.tex
\section{Background and Motivation}
\label{Sec:BackgroundAndMotivation}

In this section, we provide a brief background about probabilistic algorithms,
explain the applications structure supported by the accelerator
using motion estimation \cite{MotionEstimation} as an example, and present
the challenges for supporting uncertainty quantification and motivate our
solutions.

\subsection{Probabilistic Algorithms}
\label{SubSec:ProbabilisticAlgorithms}

Bayesian inference combines new evidence and prior beliefs to update the
probability estimate for a hypothesis. Consider $D$ as the observed data and $X$ as the
latent random variable. The prior distribution of $X$ is $p(X)$ and $p(D \mid X)$ is the
probability of observing $D$ given a certain value of $X$. In Bayesian inference, the goal
is to retrieve the posterior distribution $p(X\mid D)$ of the random variable $X$ when $D$ is
observed. As the dimensions of $D$ and $X$ grow, it often becomes difficult
or intractable to numerically derive the exact posterior distribution.

\begin{figure}
\begin{center}
\includegraphics[width=.8\linewidth,trim=0 220 0 0,clip]{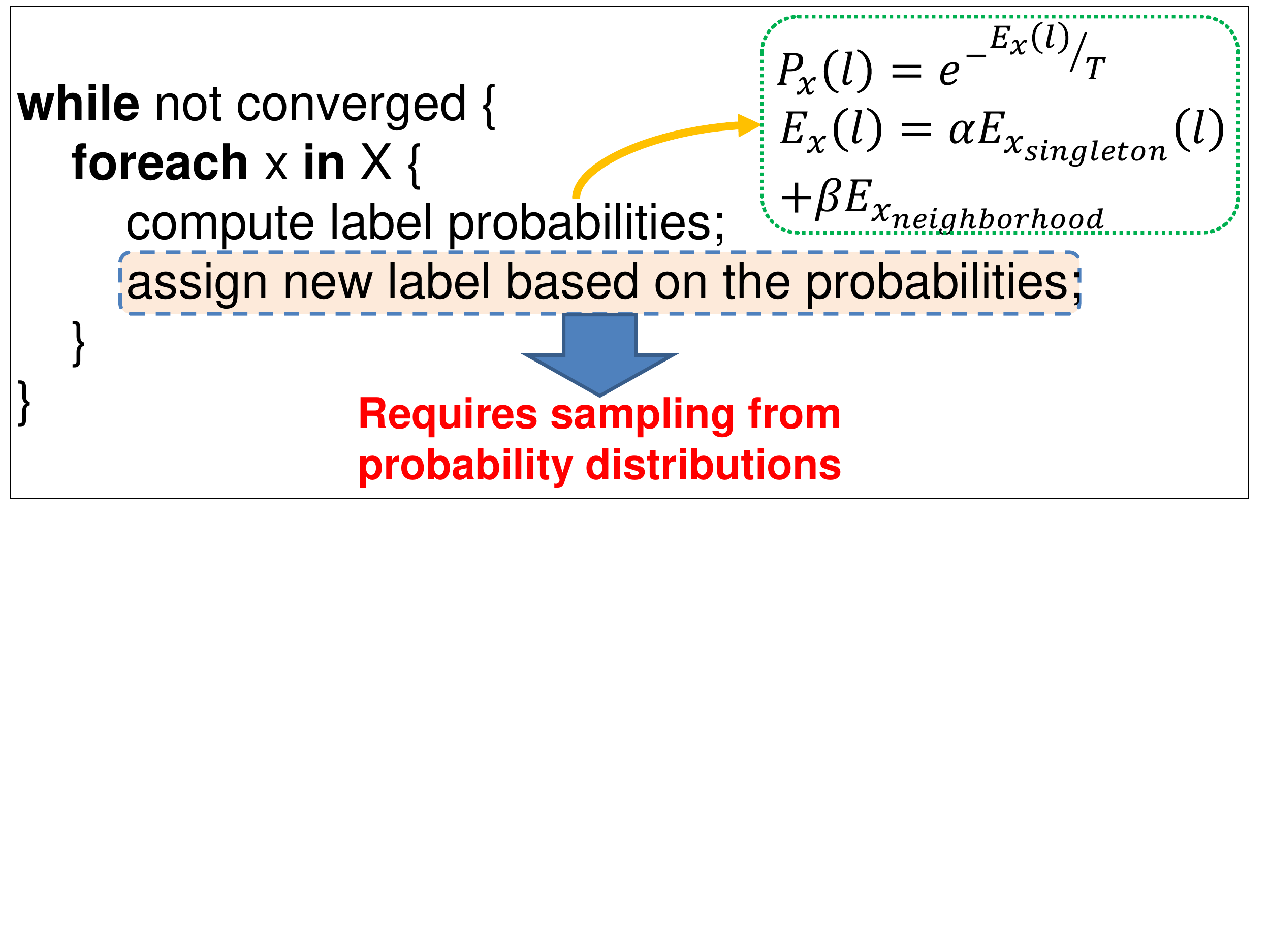}
\caption{Markov Chain Monte Carlo algorithm. Note that sampling is performed in the inner loop.}
\label{Fig:MCMC}
\end{center}
\end{figure}

One approach to solving these inference problems is to use probabilistic Markov chain Monte-Carlo
(MCMC) methods that converge to an exact solution by iteratively generating
samples for RVs (Figure \ref{Fig:MCMC}).
In practice MCMC becomes inefficient for many problems that have
high dimensionality and complex structure. It can
require many iterations before convergence, and the inner loop in
includes generating samples from
probability distributions which is computationally expensive for
conventional processors \cite{RSUG} and thus, a specialized
accelerator is needed to address these shortcomings.

\subsection{Example Application: Motion Estimation}
\label{SubSec:MotionEstimation}

\begin{figure}
\begin{center}
\includegraphics[width=\linewidth]{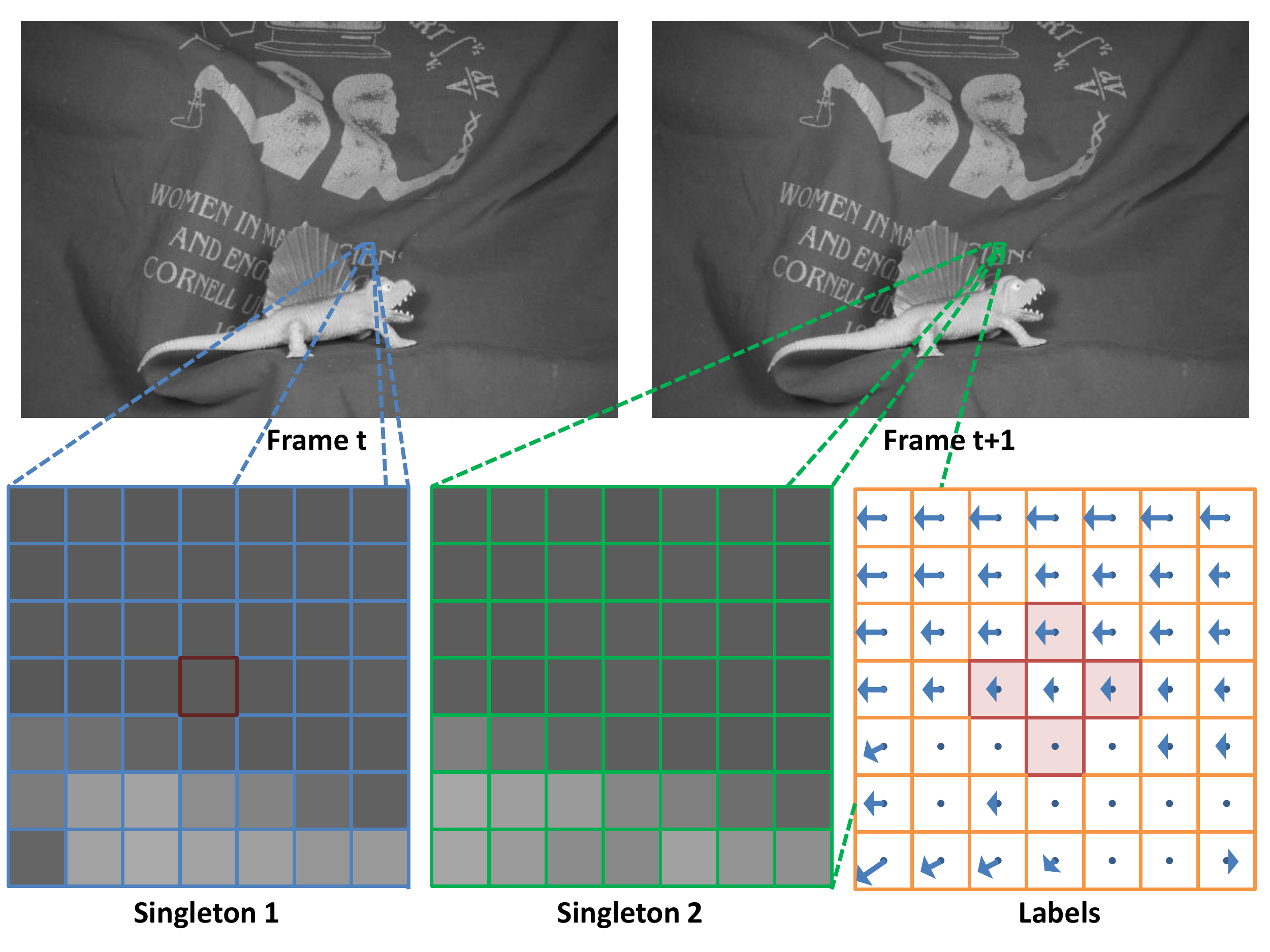}
\caption{Data access patterns for first-order MRF inference using MCMC to solve motion estimation (\textit{dimetrodon} from Middlebury database \cite{MiddleburyMotionEstimation}).}
\label{Fig:MotionEstimation}
\end{center}
\end{figure}

To shed more light on the details of the first-order MRF
inference using MCMC algorithm, we explain how first-order
MRF can be utilized to represent the motion estimation problem
and how MCMC with Gibbs sampling can solve it \cite{MotionEstimation}. The goal
is to estimate the 2-D motion vectors between two time-varying
images, such as two consecutive frames of a video.
Figure \ref{Fig:MotionEstimation} (top) shows a sample input for this problem.

In this example, the target is to compute the motion vector of the pixel in the center of the blue box
in Figure \ref{Fig:MotionEstimation} (bottom-left). To do so, the inner loop in Figure \ref{Fig:MCMC} must be executed,
which includes computing probability values according to the equations in the figure.
In the equations, $P_x(l)$ is the probability that RV $x$ takes on label
$l$,\footnote{The term \textit{label} is commonly used to indicate the value of the random variable.} $E_x(l)$ is the energy of label $l$ which depends on singleton and neighborhood
values, and $\alpha$, $\beta$, and $T$ are application parameters. $E_{x_{singleton}}$
depends on two types of singleton data: i) singleton 1 which is the gray-scale value of the pixel
itself, and ii) singleton 2 that is the gray-scale value of each of the pixels inside the green
box in Figure \ref{Fig:MotionEstimation} (bottom-middle), which form a $7\times 7$ window
surrounding the aforementioned pixel corresponding to possible labels (in this application,
each motion vector) and come from frame $t+1$. The pattern of singleton 2 is application
specific. Generally, for each RV, $E_{x_{neighborhood}}$ may depend
on the latest labels of all other RVs. However, for the first-order MRF model,
$E_{x_{neighborhood}}$ is calculated using the current labels of the top, down,
left, and right neighbors of the pixel, the shaded boxes
shown in Figure \ref{Fig:MotionEstimation} (bottom-right),
each of which are motion vectors themselves. This neighborhood pattern is fixed for the
first-order MRF model. Once the probabilities for all possible
labels are calculated, they are used to create a probability
distribution function (PDF), which in turn
is used for sampling and determining the new label for the pixel.

This process must be repeated for all pixels in frame $t$ for a certain number of iterations
(until the algorithm converges to the final solution) to obtain their motion vectors. CMOS
specialization and pseudo-random number generation can be used to accelerate the computations
required for updating each pixel. Previous work proposes a function unit for this purpose \cite{SPU}, which
is briefly reviewed in Section \ref{SubSec:SPU}. Furthermore, the structure of the MRF model
provides opportunities for parallelism (explained in Section \ref{SubSubSec:UpdateOrder}).
But there are challenges in realizing this parallelism due to the memory
access patterns which require careful memory banking and access scheduling
that are discussed in Sections \ref{SubSubSec:SingletonMemory} and \ref{SubSubSec:LabelMemory}.

\subsection{Uncertainty Quantification}
\label{SubSec:UncertaintyQuantification}

\begin{figure}
\begin{center}
\includegraphics[width=\linewidth,trim=20 160 20 160,clip]{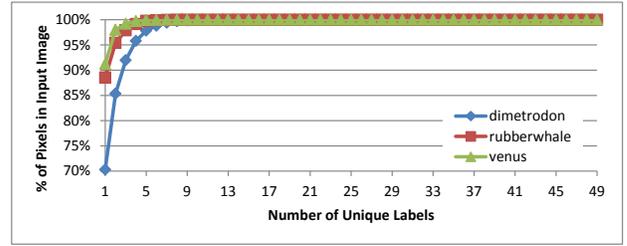}
\caption{Cumulative distribution of pixels per number of unique labels in three input datasets \cite{MiddleburyMotionEstimation} for motion estimation.}
\label{Fig:UniqueLabelsMotionEstimation}
\end{center}
\end{figure}

Probabilistic models and algorithms are “conceptually
simple, compositional, and interpretable” \cite{ProbabilisticMachineLearning}, and provide opportunity
to determine why a given result is obtained. This is due to two reasons: i) models
such as MRF inherently have transparent structures, and ii) these algorithms
allow for quantifying the uncertainty to evaluate the confidence on the obtained result.
Uncertainty quantification can be achieved by collecting a histogram of the RV's labels
after the warm-up period of the MCMC (i.e., the iterations at the beginning
of the algorithm before mixing has happened), which can then be used to
derive statistics such as mode, variance, etc., that illuminate the
uncertainty associated with the final result. In the example of image
segmentation guiding a surgeon's decision regarding tumor resection,
if the variance in the final result is high then the surgeon might
decide to remove a larger section to be safe, without removing
too much of the tissue.

However, na\"ively storing the histogram data imposes a significant memory capacity, bandwidth and
processing overhead and therefore, a more scalable solution is required for
uncertainty quantification. Fortunately, there is opportunity for optimization because
after warm-up, the RVs tend to take on only a limited number of labels.
Figure \ref{Fig:UniqueLabelsMotionEstimation} illustrates this fact.
In three input datasets \cite{MiddleburyMotionEstimation} for
motion estimation which has 49 labels, at most only
14.7\% of pixels take on more than two unique labels
during the second half of the iterations (i.e., iterations
1500-3000 in this experiment). This allows
on-chip memory space to store only more frequently
picked labels, and occasionally send the rest to
off-chip memory. Section \ref{SubSubSec:LabelMemory} presents
a hybrid on-chip/off-chip memory system for collecting the
histogram of labels based on this analysis.

%% file: 3-spa.tex
\section{Design Overview and Challenges}
\label{Sec:DesignOverviewChallenges}

The characteristics of MCMC and MRF, covered in
Sections \ref{SubSec:ProbabilisticAlgorithms} and \ref{SubSec:MotionEstimation}, guide our
design choices for the proposed accelerator. In this section,
we provide an overview of our design, the challenges
presented, and our proposed solutions.

MCMC is an iterative algorithm in which the computations of
each iteration depend on those of the previous iteration (Section \ref{SubSec:ProbabilisticAlgorithms}).
Therefore, we decide to use on-chip memory to store data and
intermediate iteration results to avoid frequent costly
off-chip communication that uses up significant bandwidth and
imposes high latencies. Furthermore, due to the structure of the
first-oder MRF, all computations are local, i.e., updating RVs
only needs data from nearby memory locations. \textit{Thus, we propose
to use a tiled architecture where each tile has its own memory
and is responsible for computations on the portion of the graphical model
stored in its memory.} This allows us to expose existing parallelism
in first-order MRF and take advantage of near-memory computing,
and eliminates the need for complex centralized coordination. The
individual tile's architecture is discussed in detail in Section \ref{SubSec:SPE}.

The proposed architecture needs a communication
infrastructure to efficiently transfer data among tiles
when needed. Particularly, due to singleton 2's
application-specific nature, designing such a communication
infrastructure 
without sacrificing flexibility can be challenging. \textit{We propose
a topology and data mapping scheme tailored to
the first-order MRF characteristics which together ensure
no communication longer than one hop},
and therefore, the overheads of a full-blown Network-on-Chip
(NoC) are avoided. Although our design is tailored to the
first-order MRF neighborhood structure, it supports arbitrary
accesses to the singleton 2 memory (S2Mem). In other words, our
proposed topology and data mapping scheme for singleton 2 do
not limit the MRF applications the accelerator can run.
Sections \ref{SubSec:SPATopology} and \ref{SubSec:Runtime} explain
the proposed network topology and data mapping schemes.

Moreover, exposing the potential parallelism inherent
in the model requires a suitable scheduling technique that
allows updating multiple conditionally independent RVs
simultaneously. We use known techniques to develop a chromatic
schedule of conditionally independent RVs that can be updated
in parallel. The implication of this scheduling technique is
that in addition to the parallelism between tiles, we can
include more than one function unit in each tile to exploit
intra-tile parallelism. However, this
introduces competing accesses to S2Mem.
Furthermore, labels memory (LMem) must be accessed at four
different locations for each RV. \textit{Therefore, we utilize
memory banking mechanisms for each of S2Mem (Section \ref{SubSubSec:SingletonMemory})
and LMem (Section \ref{SubSubSec:LabelMemory}) to facilitate stall-free execution in tiles.}

Finally, uncertainty quantification requires tracking
how many times a label is chosen for a given RV.
A na\"ive implementation requires either i) having enough
counters on the chip to keep track of all possible labels
for all RVs, which is prohibitive in terms of area, or ii)
sending the result of all label updates off chip, which
needs significant communication bandwidth.
\textit{Because of limited memory capacity on the chip,
and inspired by the insights from Figure \ref{Fig:UniqueLabelsMotionEstimation},
we design a hybrid on-chip/off-chip memory system to
store the histogram information for RVs throughout the
execution of MCMC in the form of a log.} To this end, we augment entries in
LMem with counters that keep track of how many times
each label has been picked, and only transfer this
information to an off-chip memory when it is necessary. Importantly, our proposed logging technique can be used
by any MCMC accelerator to support uncertainty quantification
with low off-chip bandwidth, low memory capacity requirements,
and low latency.
Section \ref{SubSubSec:LabelMemory} describes the design of this memory system
in more detail.

\section{Stochastic Processing Accelerator}
\label{Sec:SPA}

\begin{figure}
\begin{center}
\includegraphics[width=\linewidth,trim=0 170 0 0,clip]{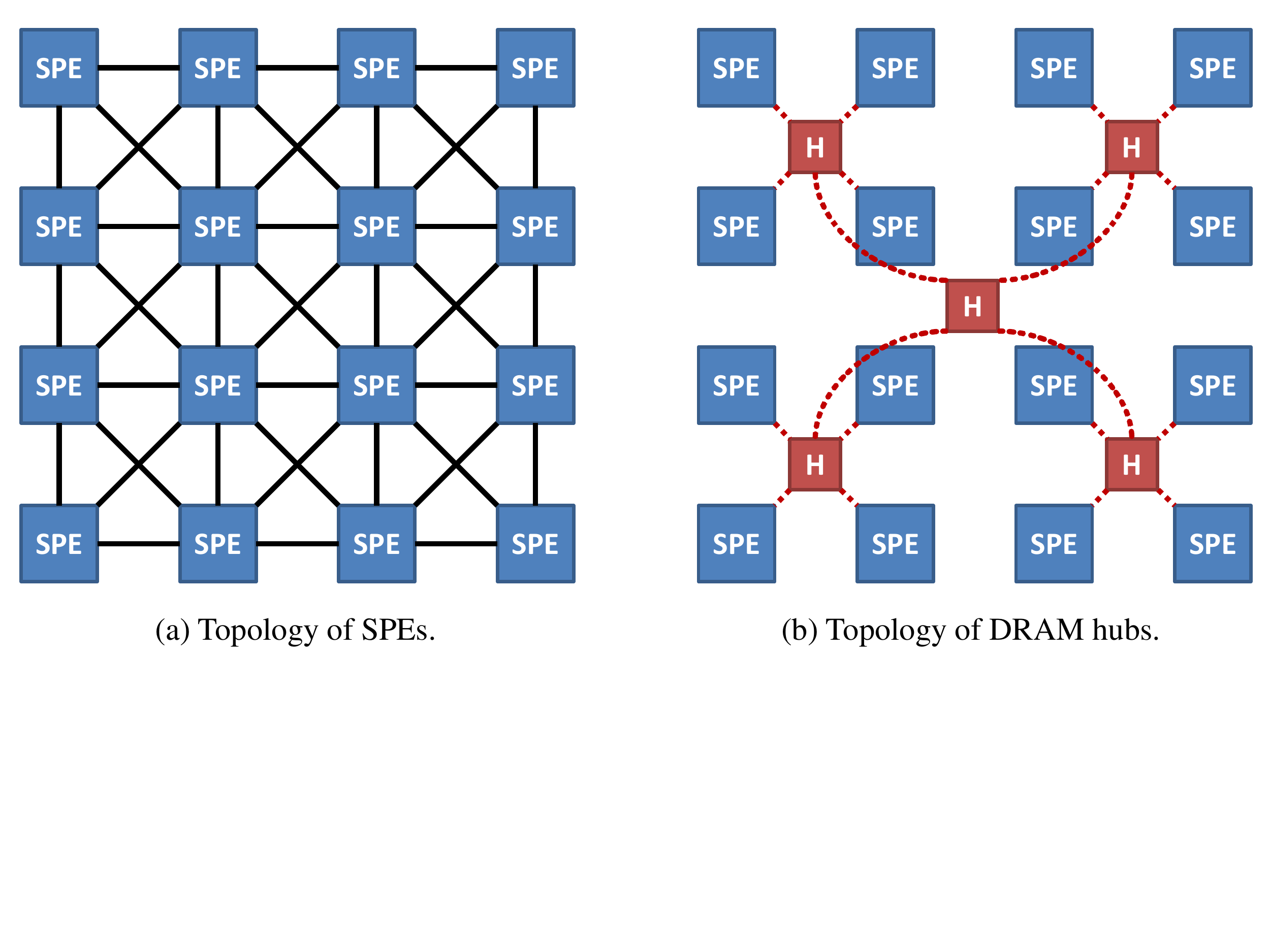}
\caption{Overview of the proposed accelerator's architecture. 
Communications between SPEs are bi-directional, but communications between
SPEs and DRAM hubs and among DRAM hubs are uni-directional. Also, the
diagonal links between SPEs are only for communicating singleton 2 data,
whereas the vertical and horizontal links transfer both label data and
singleton 2 data.}
\label{Fig:SPAOverview}
\end{center}
\end{figure}

\subsection{Overview}
\label{SubSec:SPAOverview}

Our proposed architecture for accelerating first-order MRF inference using MCMC
with Gibbs sampling is presented in this section. Figure \ref{Fig:SPAOverview}
shows an overview of the accelerator's architecture. It is composed of a
number of computation tiles or SPEs (Section \ref{SubSec:SPE}),
and DRAM hubs that route communication to the off-chip DRAM.
Each SPE is responsible for processing a portion of the
input to take advantage of near-memory computing and exploit the inherent
parallelism of the model. It comprises a number of Stochastic Processing
Units (SPUs) \cite{SPU}, which perform the main MCMC computations (Section \ref{SubSec:SPU}),
in addition to a scheduler which sequences through RVs (Section \ref{SubSubSec:UpdateOrder}),
a portion of the singleton memories (Section \ref{SubSubSec:SingletonMemory}) and the label
memory (Section \ref{SubSubSec:LabelMemory}), on which it performs the MCMC
updates, and communication components that transfer data between different SPEs
and between the accelerator and the off-chip DRAM that stores the histogram log
of the labels. Each SPE is connected to all its nearest SPEs and only communicates
with those (Section \ref{SubSec:SPATopology}). To ensure that communications
longer than one hop are never required, appropriate data mapping and
data replication schemes are adopted which are handled by the runtime (Section \ref{SubSec:Runtime}).

\subsection{Stochastic Processing Unit}
\label{SubSec:SPU}

\begin{figure}
\begin{center}
\includegraphics[width=\linewidth,trim=0 280 50 0,clip]{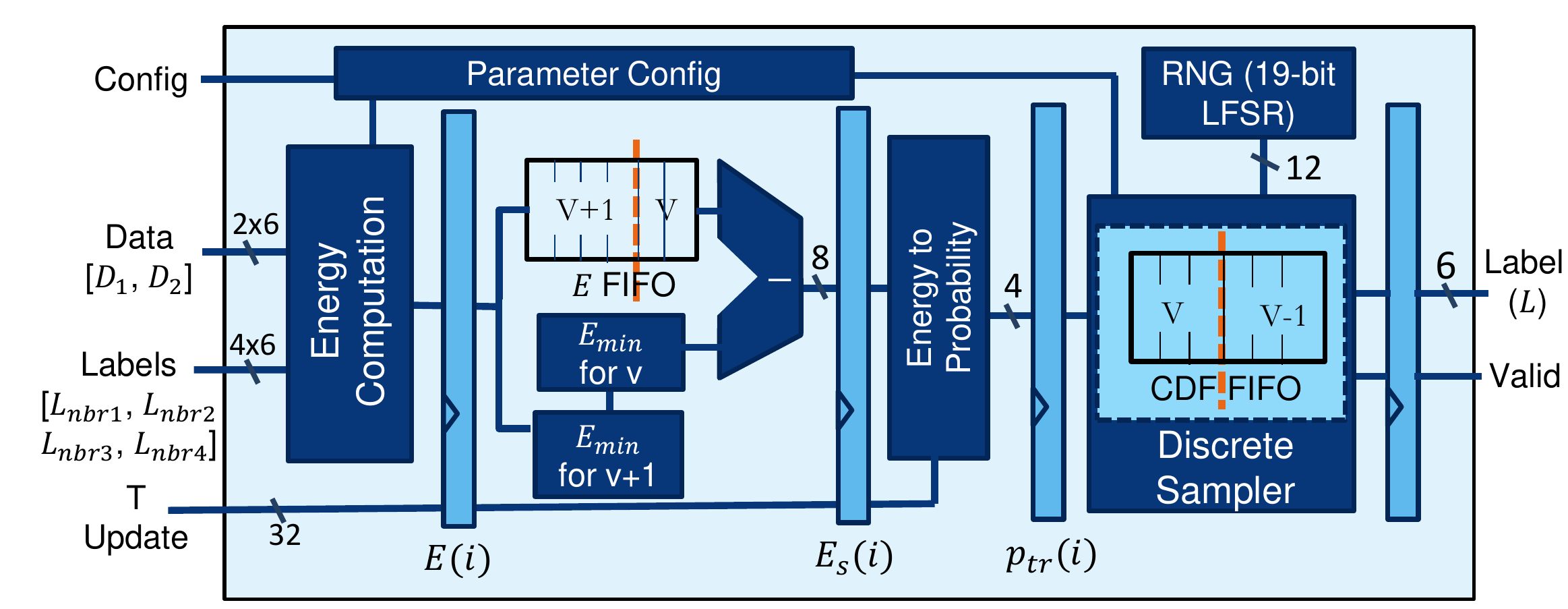}
\caption{The SPU microarchitecture reproduced from \cite{SPU}.}
\label{Fig:SPU}
\end{center}
\end{figure}

Zhang et al. propose a Gibbs sampling function unit, called Stochastic
Processing Unit (SPU), that utilizes specialization and pseudo-random number
generation to accelerate MCMC computations \cite{SPU}. Figure \ref{Fig:SPU} demonstrates
the microarchitecture of this function unit. It is composed of four main
pipeline stages, namely energy computation (Equation \ref{Eq:Energy}), dynamic
energy scaling (Equation \ref{Eq:DynamicScaling}), energy to probability conversion
(Equations \ref{Eq:ScaledProb} and \ref{Eq:TruncatedProb}), and sampling.

\begin{equation} \label{Eq:Energy}
E(l) = \alpha E_{singleton}(l) + \beta \sum{E_{neighborhood}}
\end{equation}
\begin{equation} \label{Eq:DynamicScaling}
E_s(l) = E(l) - E_{min}
\end{equation}
\begin{equation} \label{Eq:ScaledProb}
P_s(l) = (2^{P_{bits}}-1)\times exp(-E_s(l)/T)
\end{equation}
\begin{equation} \label{Eq:TruncatedProb}
P_{tr}(l) = \lfloor 2^{\lfloor \log_2 P_s(l) \rfloor} \rfloor 
\end{equation}

Energy computation takes the singleton data and neighbor labels, all 6-bit values,
and computes the energy of a possible label, $E(l)$ in Equation \ref{Eq:Energy},
where $\alpha$ and $\beta$ are application parameters. Next, $E(l)$
is dynamically scaled by subtracting the minimum energy of all labels
from it to maximize the dynamic range. Energy values (raw and scaled)
are 8-bit unsigned integers. The scaled energy $E_s(l)$ is then converted
to a scaled probability represented by a 4-bit unsigned integer. The original
probability (real number in $[0,1]$) is calculated using $exp(-E_s(l)/T)$,
in which $T$ is a per iteration parameter. To avoid using floating-point function
units, though, the probability is scaled using Equation \ref{Eq:ScaledProb},
and then truncated using Equation \ref{Eq:TruncatedProb}. $P_{bits}=4$ ensures
the scaled probability is in $[0,16]$, which allows for representing the
number using 4 bits. Afterward, Equation \ref{Eq:TruncatedProb} approximates
the scaled probabilities to the nearest power of two, i.e., $P_{tr}\in\{0, 1, 2, 4, 8\}$.
The possible values of $P_{tr}(l)$ can be pre-computed
and stored in a look-up table (LUT). These values must be updated if $T$ changes.
The last stage generates a sample per RV based on $\{P_{tr}(0), P_{tr}(1), ..., P_{tr}(L-1)\}$,
where $L$ is the number of labels, using the least significant twelve bits
of a 19-bit Linear Feedback Shift Register (LFSR) to implement the inverse
transform sampling. The SPU's throughput is one RV update per $L$ cycles,
if it receives the appropriate input (i.e., neighborhood labels and singleton
data) at every cycle. Our goal in Sections \ref{SubSec:SPE} and \ref{SubSec:SPATopology}
is to design an architecture that ensures this condition is realized.

The SPU can be used in one of two modes: i) pure sampling, or ii) optimization.
The main difference between these two modes is that in pure sampling, the
parameter $T$ is the same for all Gibbs sampling iterations, whereas in
optimization (simulated annealing), $T$ gradually decreases to help faster
convergence to a final solution \cite{Gibbs}. The implication of this difference is that
in pure sampling mode, when the algorithm converges to a final solution,
the estimated distribution of a RV can be generated by collecting the
histogram of the latest N samples. Since in this work we are interested
in the uncertainty quantification capability of MCMC, we only focus on
the pure sampling mode. However, the proposed accelerator can operate
in optimization mode as well.

\subsection{Stochastic Processing Element}
\label{SubSec:SPE}

\begin{figure}
\begin{center}
\includegraphics[width=.8\linewidth,trim=10 75 220 20,clip]{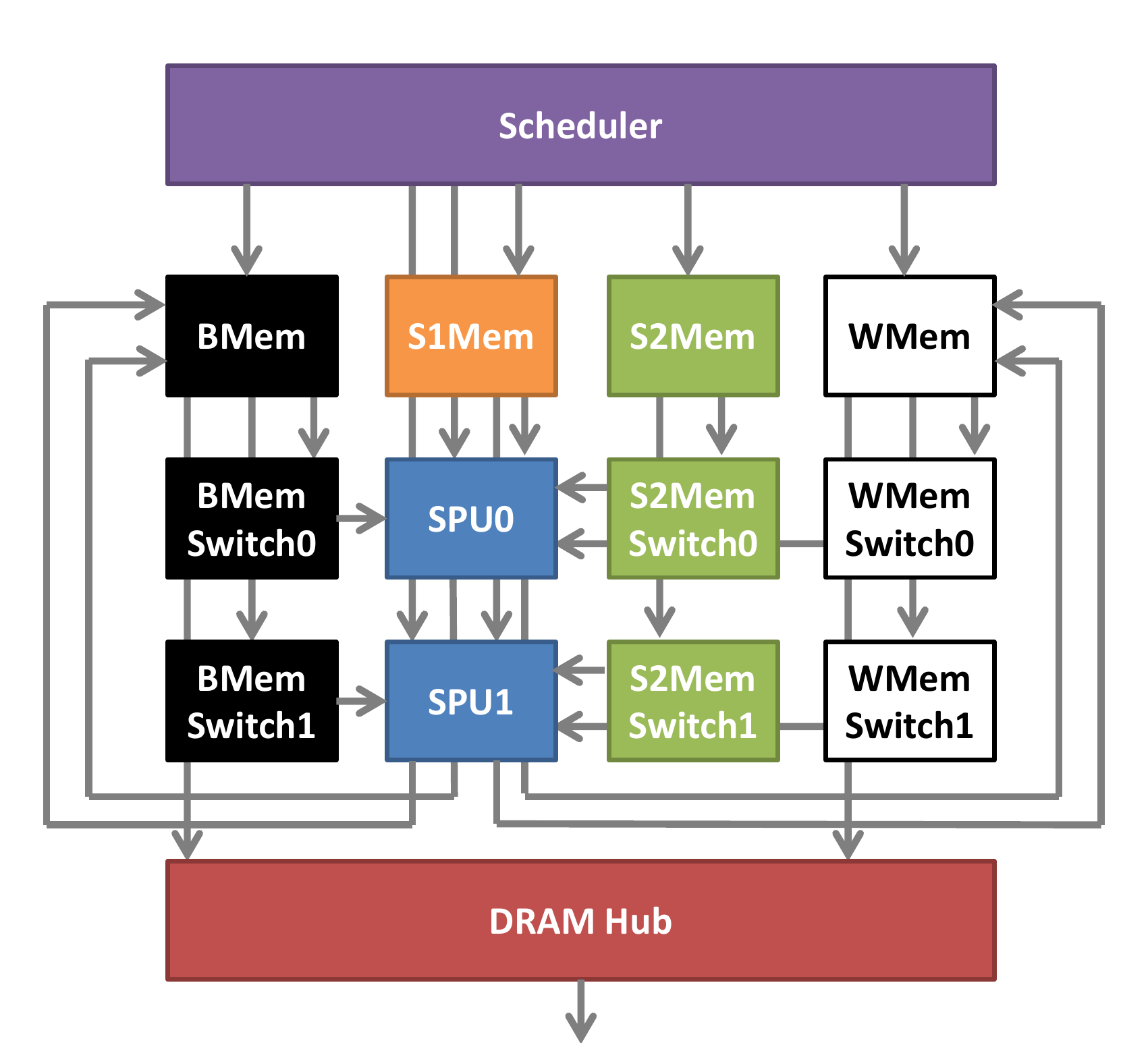}
\caption{Architecture of an SPE with two SPUs. For the sake of clarity,
non-local outgoing communications from BMem, WMem, and S2Mem modules and incoming
communications to BMem Switches, WMem Switches, and S2Mem Switches have been omitted.
BMem and WMem send data to the BMem and WMem Switches in the top, down,
left, and right neighbors. S2Mem sends data to the S2Mem Switches in all
the eight neighbors shown in Figure \ref{Fig:SPAOverview}a.}
\label{Fig:SPE}
\end{center}
\end{figure}

An SPE incorporates the components that carry out the operations needed to
feed the necessary data to SPUs every cycle and write back the result
of their computations to the LMem. These operations include
sequencing through RVs and generating memory addresses
corresponding to the singleton and neighborhood data, reading the
data at those addresses and passing them to the appropriate SPU,
and writing the results of the computations back to the correct
addresses in LMem. Furthermore, to maintain
the histogram of labels, it might be necessary that while
writing data back to LMem, some data be sent to
the off-chip DRAM. Figure \ref{Fig:SPE} shows the SPE's components
and the interactions between them. The next section
describes the various components inside
an SPE, followed by a more detailed explanation of scheduling and
different types of memory in the SPE.

\subsubsection{Components} ~
\label{SubSubSec:SPEComponents}

\textbf{Scheduler:} This component's main job is to generate the
update schedule and coordinate the operations of most of the other
components in an SPE. It interacts with SPUs and various memory
blocks, and its functionalities include sending the SPUs some
parameters including the $T$ in MCMC equations in Figure \ref{Fig:MCMC},
and other information such as whether a pixel is on the boundary
or whether it is a black or a white pixel (to determine the
destination of the computation results). It also sends computed
addresses to different memory blocks, so that they can return the
requested data to the SPU.

\textbf{Singleton 1 Memory:} Denoted by S1Mem in Figure \ref{Fig:SPE},
it stores singleton 1 data as the name suggests, or in the example
of motion estimation in Section \ref{SubSec:MotionEstimation}, the
data in the blue box in Figure \ref{Fig:MotionEstimation}. It receives
addresses from the Scheduler and sends data to the SPUs once for
every RV.

\textbf{Singleton 2 Memory:} Similar to S1Mem, it is
referred to as S2Mem in Figure \ref{Fig:SPE}, and stores singleton 2
data (i.e., the data in the green box in Figure \ref{Fig:MotionEstimation}).
Because each singleton 2 corresponds to an individual label, as opposed to
singleton 1 which is fixed for all labels of the same RV, S2Mem
receives a base address from the Scheduler, and computes addresses
for the appropriate singleton 2 data point for each label by reading from
an offset look-up table (LUT) populated by the runtime in an application-by-application
basis. For instance, in the case of motion estimation, the LUT stores
the offsets that define the $7\times 7$ window shown in Figure \ref{Fig:MotionEstimation}.
It then sends that data to S2Mem Switches for every label. Since
this data is needed for every label at every SPU, it is required that
multiple reads from different addresses be issued at the same cycle.
We address this problem by devising a banking scheme that is described
in Section \ref{SubSubSec:SingletonMemory}. Both singleton memories are read-only,
meaning they get initialized in the beginning by
the runtime and will never change throughout the execution.

\textbf{Singleton 2 Switch:} These switches receive data from the
appropriate S2Mem, i.e., either the local S2Mem or one of
the memories in one of the eight neighbors, and
send it to the SPUs they are connected to.

\textbf{Label Memories:} BMem and WMem in Figure \ref{Fig:SPE},
toghether form the LMem. These memories store the results of the computations done by the
SPUs. They receive addresses from the Scheduler to send neighborhood
data to the corresponding switches, which in turn send those data to
the SPUs. They also receive the new labels from SPUs. Although neighborhood
data is needed only once per RV, due to the model's
structure multiple reads must be issued simultaneously to provide
the data necessary for beginning the computations to the SPUs.
We solve this problem by banking the LMem and pipelining
accesses to them. In addition to storing the labels computed
by the SPUs, LMem is also part of the hybrid on-chip/off-chip
memory system that stores the information required for generating
the labels histogram. LMem is explained in more detail in Section \ref{SubSubSec:LabelMemory}.

\textbf{Label Switches:} These switches are similar in functionality
to S2Mem Switches, i.e., they receive neighborhood data from
the local LMem as well as the LMems in the top,
down, left, and right SPEs and pass them to their corresponding SPU.

\subsubsection{Updating Order and Inter-variable Parallelism} ~
\label{SubSubSec:UpdateOrder}

In general, MCMC is a sequential algorithm since updating each RV
depends on the latest value of all other RVs. However,
as explained in Section \ref{SubSec:MotionEstimation}, in the
first-order MRF model, each RV is only conditionally
dependent on its top, down, left, and right neighbors. This means
there is opportunity to develop a chromatic schedule for
updating conditionally independent variables in parallel and thus,
significantly reduce the execution timem. For first-order MRF,
this schedule is a simple checkerboard
scheme which divides the random field into a black (BMem) and a white (WMem)
subset, where all RVs in each subset are independent \cite{ParallelGibbs, EricJonas, FPGAGibbs}.

In our proposed accelerator, the Scheduler component in each SPE
is responsible for generating this schedule. The Scheduler first
goes through all black RVs, then flushes the pipeline
of all other components, and repeats the same process for all
white RVs.

Another benefit of this chromatic schedule is that it puts restrictions
on access types to different parts of LMem, i.e., while
black variables are being updated, there will be no writes issued to
WMem and vice versa. This allows
for simplifying the memory structure by dividing it into a black and
a white region, knowing that the Scheduler takes care of avoiding
conflicting accesses to these regions.

\subsubsection{Singleton Memory Structure for Multiple SPUs} ~
\label{SubSubSec:SingletonMemory}

\begin{figure}
\begin{center}
\includegraphics[width=\linewidth,trim=0 265 0 0,clip]{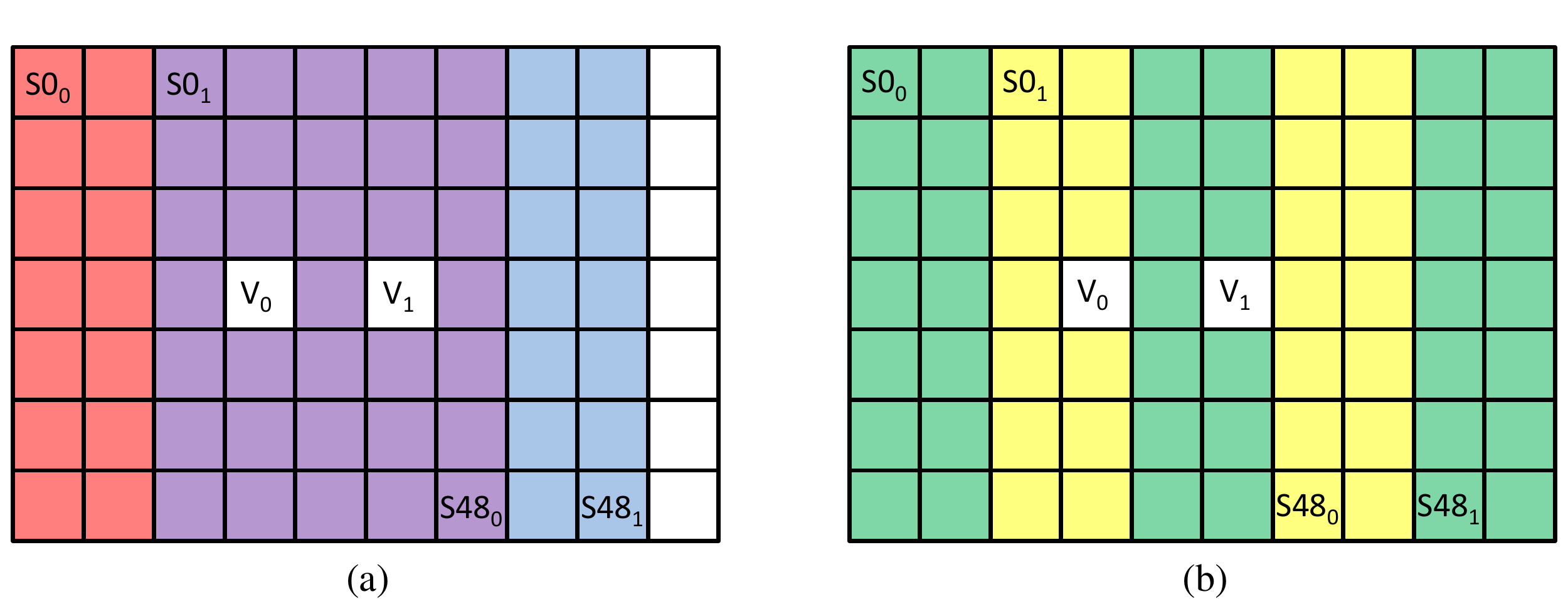}
\caption{(a) An example of the singleton 2 access pattern for motion estimation in an SPE with two SPUs. Boxes shaded in red and purple illustrate the singleton 2 window for RV $V_0$, and boxes shaded in purple and blue show that window for $V_1$, and (b) the proposed banking scheme to solve the multiple simultaneous accesses issue. Each color represents one bank. Locations denoted by $S0_0$ and $S0_1$ must be accessed together, and so is the case for $S48_0$ and $S48_1$.}
\label{Fig:Singleton2Example}
\end{center}
\end{figure}

According to the MCMC details explained in Section \ref{SubSec:MotionEstimation}, there are
potentially two types of singleton data, both of which are stored
in read-only memories: 1) singleton 1, which is always present and is required once for
a RV, and 2) singleton 2, which, if present, is required
for each possible label a RV can take on. The implication of
singleton 2's access pattern is that if we decide to have
more than one SPU in an SPE
to amortize the area of the Scheduler and other control logic, servicing
singleton 2 reads becomes a challenge. Figure \ref{Fig:Singleton2Example}a
illustrates this problem with an example of two
SPUs in an SPE running in parallel. Multiple
pieces of singleton 2 data at different addresses must be read
at the same cycle. There are three possible solutions to
accommodate this access pattern:

\begin{enumerate}
    \item S2Mem must support a read size larger than one singleton 2, and an intermediate register must handle the feeding of data to the appropriate SPU. This option also allows exploiting the temporal locality, i.e., a piece of data can be read once and used multiple times if it is required for multiple RVs. However, determining when to issue new reads, shifting and moving data around, and developing an update schedule that matches this design make it complicated, particularly due to the application-specific patterns of singleton 2 accesses.
    \item S2Mem must be a multi-port structure to straightforwardly read the required data from it. Nevertheless, multi-port memories are area- and power-hungry and are generally not preferable \cite{CMOSVLSI}. This option also does not take advantage of singleton 2's temporal locality.
    \item S2Mem must be divided into separate banks which have only one port and are accessed simultaneously. Similar to the previous solution, the drawback of this design compared to the first one is that it too necessitates reading the same piece of data multiple times. However, it allows for a simpler Scheduler and memory structure and therefore, we choose this option for S2Mem.
\end{enumerate}

Our proposed banking scheme exploits the knowledge of the
update order discussed in Section \ref{SubSubSec:UpdateOrder}. More
specifically, we take advantage of the stride of two consecutive RVs
in the same row. Because we know the next RV will always
be two locations ahead and the singleton 2 access pattern is
the same for all RVs, it logically follows that the nest singleton 2
will also be two locations ahead. Thus, we put every two
columns of singleton 2 in a separate bank, for a total number
of banks equal to the number of SPUs inside the SPE. Figure \ref{Fig:Singleton2Example}b
demonstrates this for an example SPE with two SPUs. The runtime is
responsible for correctly populating these banks.

\subsubsection{Labels Memory and Labels Log} ~
\label{SubSubSec:LabelMemory}

\begin{figure}
\begin{center}
\includegraphics[width=\linewidth,trim=20 380 20 0,clip]{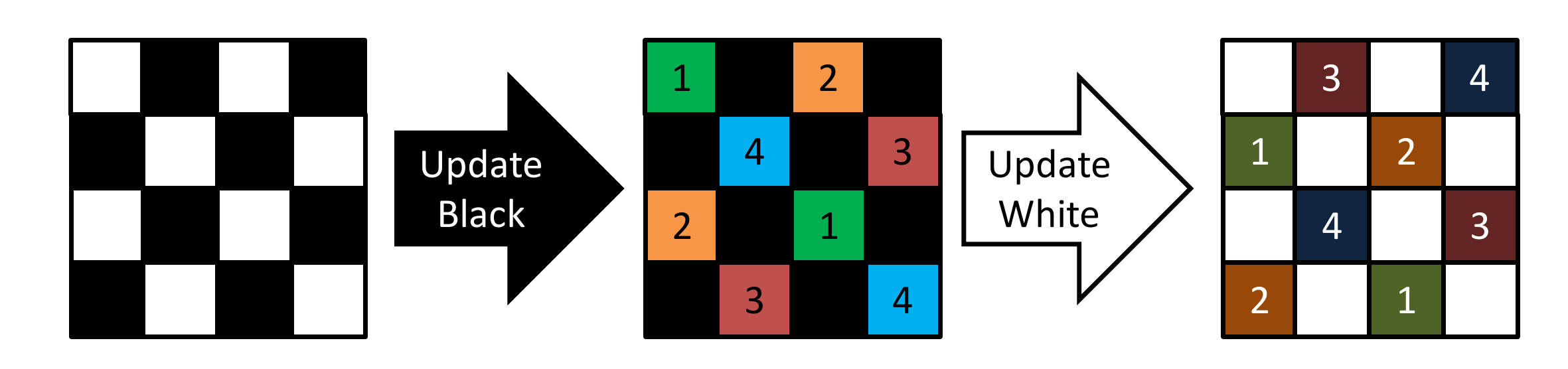}
\caption{Memory banking scheme for white and black sections of LMem. Note that while updating RVs in BMem, their neighbors are in WMem and vice versa.}
\label{Fig:LabelMemBanking}
\end{center}
\end{figure}

Similar to singleton 1, neighborhood data is needed once
for each RV (Section \ref{SubSec:MotionEstimation}).
Nevertheless, since in the first-order MRF the neighborhood
structure consists of four RVs, a simple monolithic single-port
memory does not accommodate the requirements of the
model. A key difference with S2Mem, though,
is that neighborhood data is needed only once for each
RV. Therefore, reads to the LMem can be pipelined
to provide data to multiple SPUs in an SPE. To overcome
the challenge of accessing four locations in
LMem simultaneously, we use a memory
banking scheme shown in Figure \ref{Fig:LabelMemBanking}.
This pattern is repeated to cover all the RVs in the
input. It ensures that for every RV, its top, down,
left, and right neighbors reside in unique banks.

In addition to storing the labels data, LMem is also a part
of the on-chip/off-chip memory system that collects the histogram of
labels for uncertainty quantification. Collecting an accurate histogram needs a
counter per each possible label, which in our case would be 64, and
each counter must be able to hold a maximum value of the maximum number
of iterations, which could be 10-12 bits. Keeping such a huge
amount of data on chip is neither practical nor efficient. Fortunately,
it is not necessary either.

As previously shown (Section \ref{SubSec:UncertaintyQuantification}, Figure \ref{Fig:UniqueLabelsMotionEstimation}), a significant portion
of RVs take on only a few unique labels after the warp-up
period has passed. This inspired us to have room for a few labels and
their corresponding counters on chip, and once a counter is saturated
or a new label is selected that is not present in the LMem,
send a message consisting the evicted label's ID, the RV's
address, and the count associated with it, to an off-chip memory. This
data is stored in the form of a log, which at the end of the execution
is processed by the runtime and translated into a histogram with no loss
in accuracy, because all the label count information is preserved in the off-chip log.
The operation of this memory structure is similar to a write-back,
write-allocate, no fetch-on-write cache. The main
advantage of such a design is that unlike normal caches where data
travels in both directions (i.e., on-chip to off-chip and vice versa),
here data only go out from on-chip memory and hence,
with deep enough FIFOs to store the messages until they can be sent
to the off-chip memory, the computation units will not be forced to stall.

The remaining challenges are: 1) choosing an efficient replacement policy,
and 2) determining the optimal size of the on-chip LMem (i.e., how many
label+counter pairs to keep per RV). We considered two
replacement policies, Least Frequently Picked (LFP), and Least Recently
Picked (LRP). Intuitively, LFP makes the most sense because we want to
keep the label that is selected most often on-chip. However, it is both
more complicated to implement, and more sensitive to the time we start
to collect the histogram. To be more specific, if we start collecting
the histogram too soon, i.e., before the end of the warm-up period, it is
possible that a label which is not among the top few most frequently
picked labels overall is picked enough times that it prevents actual frequent
labels from remaining in the on-chip memory. LRP, however, avoids this
by evicting the aforementioned label because it is not selected
anymore after the warm-up period. For these reasons, we choose LRP
as the replacement policy.

To determine the size of the on-chip memory, we must take into account
the trade-off between this size, and the off-chip bandwidth and the size
of the off-chip log. Ideally, we want the smallest on-chip
memory that the off-chip bandwidth allows. We use Equation \ref{Eq:HistogramBandwidth}
to arrive at this size:

\begin{equation}
    \frac{\frac{\#SPUs}{\#Labels}*EvictionRate * MessageSize}{Bandwidth} < 1
    \label{Eq:HistogramBandwidth}
\end{equation}

\begin{figure}
\begin{center}
\includegraphics[width=\linewidth,trim=20 290 20 290,clip]{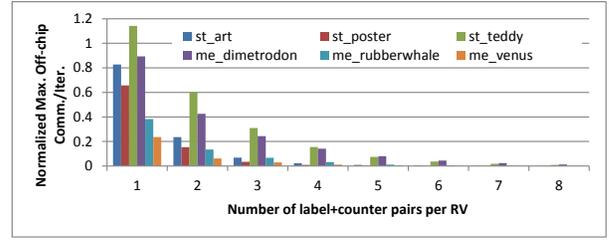}
\caption{The maximum values of Equation \ref{Eq:HistogramBandwidth} for three input data sets for stereo vision \cite{StereoVision} and motion estimation \cite{MiddleburyMotionEstimation} per LMem size. The replacement policy is least recently picked.}
\label{Fig:OffchipBandwidth}
\end{center}
\end{figure}

\begin{table}[t]
  \centering
  \caption{Values used to calculate Equation \ref{Eq:HistogramBandwidth} for Figure \ref{Fig:OffchipBandwidth}.}
  \label{Table:EquationParams}
  \begin{tabular}{|l|c|}
    \hline
    \textbf{Parameter} & \textbf{Value} \\
    \hline
    \hline
    \#SPUs & 2048\\
    \hline
    \#Labels & Stereo Vision: 28, 30, 56; Motion Estimation: 49\\
    \hline
    Message Size & 32 bits\\
    \hline
    Bandwidth & 512 bits/cycle\\
    \hline
  \end{tabular}
\end{table}

$\#SPUs$ is the total number of SPUs in the accelerator, $\#Labels$ is the
number of possible labels a RV can take on (an application-specific
value), $EvictionRate$ is the rate at which labels are evicted to off-chip
memory, $MessageSize$ is the size of the messages in bits, and $Bandwidth$ is
the available off-chip bandwidth. Equation \ref{Eq:HistogramBandwidth} indicates
that the amount of off-chip communication must not exceed the available
bandwidth. Figure \ref{Fig:OffchipBandwidth} shows the maximum value of
Equation \ref{Eq:HistogramBandwidth} for three input data sets for
stereo vision \cite{StereoVision} and motion estimation \cite{MiddleburyMotionEstimation} for
different sizes of LMem (Table \ref{Table:EquationParams} lists the values used to compute the
result of Equation \ref{Eq:HistogramBandwidth}). To generate this graph, we first collect a
trace of the labels of all RVs at every iteration. We then
process this trace to simulate the behavior of our proposed LMem
with sizes of 1-8 label+counter pairs per RV. The figure indicates
that with a LMem large enough to hold only two label+counter pairs per RV,
the off-chip bandwidth utilization will not exceed 60\% of the available bandwidth.
Therefore, we select two label+counter pairs per RV as the size of the on-chip LMem.

\begin{figure}
\begin{center}
\includegraphics[width=\linewidth,trim=0 325 0 0,clip]{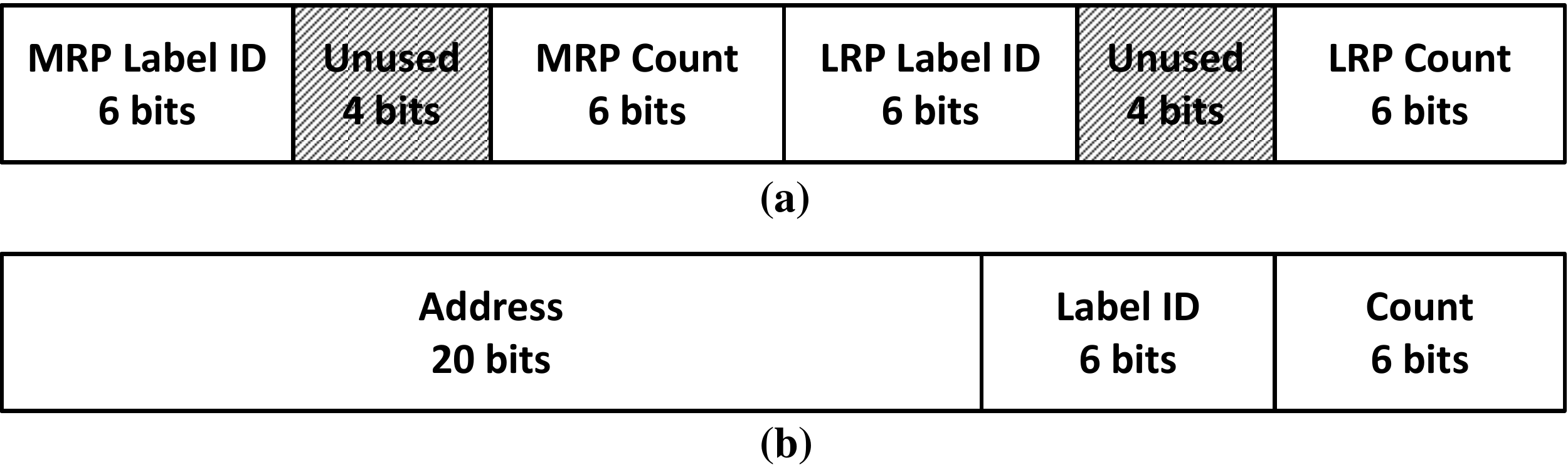}
\caption{(a) Structure of a LMem entry, and (b) structure of a message to off-chip memory, in which
address of the RV and label ID are together the bin identifier in the histogram. For the structure of the
LMem entry, we assumed that data must be stored is 32-bit words due to FPGA limitations, but more area
savings are possible by eliminating the unused bits in an ASIC implementation.}
\label{Fig:BitStructure}
\end{center}
\end{figure}

Figure \ref{Fig:BitStructure} shows the structure of a
LMem entry and a message sent to off-chip memory, given
the number of label+counter pairs per RV, and the number of RVs
that the accelerator supports (1M RVs in this example).
It is possible to change the width of the counter and the
address fields depending on the size of the target input sets.
It is also possible to reduce the required bits for addressing
by directing messages from certain SPEs to pre-defined offsets
in the off-chip DRAM.

With this proposed scheme, writing to the LMem is transformed
to a read-modify-write, where depending on the current
labels in the target LMem entry and the new label, a re-ordering of the data in the
entry or sending a message to the off-chip memory may be needed. Algorithm
\ref{Alg:LabelMEMWR} demonstrates this operation in more detail.

\begin{algorithm}
\SetAlgoLined
 \KwIn{addr, new\_lbl}
 \{mrp\_lbl, mrp\_cnt, lrp\_lbl, lrp\_cnt\} $\gets$ mem[addr]\;
 \uIf{new\_lbl = mrp\_lbl}{
  \eIf{mrp\_cnt = MAX\_VALUE}{
   \{addr, mrp\_lbl, MAX\_VALUE\} $\rightarrow$ DRAM\;
   mem[addr] $\gets$ \{mrp\_lbl, 1, lrp\_lbl, lrp\_cnt\}\;
  }{
   mem[addr]~$\gets$~\{mrp\_lbl,~mrp\_cnt~+~1,~lrp\_lbl,~lrp\_cnt\}\;
  }
}
 \uElseIf{new\_lbl = lrp\_lbl}{
  \eIf{lrp\_cnt = MAX\_VALUE}{
   \{addr, lrp\_lbl, MAX\_VALUE\} $\rightarrow$ DRAM\;
   mem[addr] $\gets$ \{lrp\_lbl, 1, mrp\_lbl, mrp\_cnt\}\;
  }{
   mem[addr]~$\gets$~\{lrp\_lbl,~lrp\_cnt~+~1,~mrp\_lbl,~mrp\_cnt\}\;
 }
 }
 \Else{
  \{addr, lrp\_lbl, lrp\_cnt\} $\rightarrow$ DRAM\;
  mem[addr] $\gets$ \{new\_lbl, 1, mrp\_lbl, mrp\_cnt\}\;
 }
 \caption{Read-modify-write operation to the label memory.}
\label{Alg:LabelMEMWR}
\end{algorithm}

\subsection{Accelerator Topology}
\label{SubSec:SPATopology}

There are two networks in our proposed accelerator, one that
connects the SPEs which transfers label and singleton 2 data
(Section \ref{SubSubSec:SPENetwork}), and another that connects the label
memories in SPEs to the interface to off-chip memory (Section \ref{SubSubSec:DRAMHubNetwork}).
These two networks carry traffic with different characteristics
and requirements, and thus, have different topologies which are 
discussed in the remainder of this section.

\subsubsection{SPE Network} ~
\label{SubSubSec:SPENetwork}

\begin{figure}
\begin{center}
\includegraphics[width=\linewidth,trim=20 160 20 160,clip]{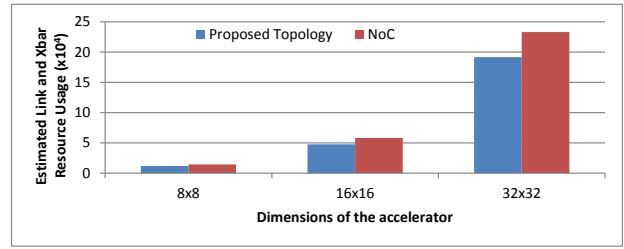}
\caption{Comparison of the estimated amount of resources needed for links and crossbars in the proposed topology and a 2-D mesh NoC.}
\label{Fig:TopologyComparison}
\end{center}
\end{figure}

Communications among SPEs follow a regular pattern, e.g., when
an SPU inside an SPE needs a piece of data that resides in the
left neighbor, the right neighbor also needs a piece of data at that
same address in the current SPE. This is true for both label data and
singleton 2 data communication, and is due to the characteristics
of the first-order MRF model and the even distribution of work to SPEs
guaranteed by the runtime.

Due to this regular pattern of communication, we propose to use
a topology in which every SPE is connected to its top, down, left,
and right SPEs for transferring label data, and to all eight
nearest neighbors (as shown in Figure \ref{Fig:SPAOverview}a) to communicate
singleton 2 data. The runtime then ensures that all the data an
SPE could possibly need reside in those SPEs to which it is directly
connected. This way, there is no need for a full-fledged
Network-on-Chip (NoC). Whenever SPEs need data from their neighbors,
they also push data in the opposite direction because their neighbor
needs the same type of data. This ensures a stall-free execution.
Additionally, this topology avoids the area overhead of a NoC router. Nevertheless,
crossbars and links are still required for moving data around. Figure \ref{Fig:TopologyComparison}
demonstrates the estimated amount of resources needed for our proposed
topology compared to a 2-D mesh NoC, for accelerators with three different
dimensions in which each SPE has 2 SPUs. The values are derived from Equations
\ref{Eq:NNLinks}, \ref{Eq:NNXbar}, \ref{Eq:NoCLinks}, and \ref{Eq:NoCXbar},
in which $D$ denotes the dimension of the accelerator, $S$ shows
the number of SPUs per SPE, $NN$ refers to our proposed topology, and $NoC$
indicates the 2-D mesh NoC. Also, $I:O$ means a crossbor with $I$ input and
$O$ output ports. To estimate the amount of resources needed for a crossbar,
we simply multiplied its number of input and output ports.
We substituted $S$ with 2 and added the two values for each
topology to generate Figure \ref{Fig:TopologyComparison}. Although this is not an accurate
measure of required resources as, for instance, one could argue
that links and crossbars should not have the same weight,
it provides a reasonable estimate. Given the estimated decreased resource usage
combined with the reduced design complexity enabled by our
proposed topology, we choose this topology over a generic NoC.

\begin{equation}
    NN_{Links} = 2 (2 (D - 1) D (S + 1) + 2 (D - 1)^2 S)
    \label{Eq:NNLinks}
\end{equation}
\begin{equation}
    NN_{XB} = D^2 (2 (4:8) + (S:9S) + 8 (2:S) + (9S:S))
    \label{Eq:NNXbar}
\end{equation}
\begin{equation}
    NoC_{Links} = 2 (2 (D - 1) D (S + 1))
    \label{Eq:NoCLinks}
\end{equation}
\begin{equation}
    NoC_{XB} = D^2 (2 (4:8) + (S:5S) + 8 (2:S) + (5S:5S))
    \label{Eq:NoCXbar}
\end{equation}

\subsubsection{DRAM Hub Network} ~
\label{SubSubSec:DRAMHubNetwork}

Unlike the regular communications between SPEs which depending
on the application can be intensive during some periods of execution,
communications between SPEs and DRAM Hubs are irregular and
designed to be infrequent. Although we cannot guarantee the
latter is always the case, our workload characterization discussed
in Section \ref{SubSubSec:LabelMemory} demonstrated that by
carefully designing the memory system, we can achieve this in
practice. Guided by this assumption, we use a tree topology for
the DRAM Hub network, as shown in Figure \ref{Fig:SPAOverview}b. Every
four SPEs are connected to one DRAM Hub, forming a region, and then
every four DRAM Hubs are connected to each other. This pattern
continues up until the interface with the off-chip DRAM. This
topology is scalable and does not cause communication with the DRAM
to become a bottleneck. Furthermore, communication with the DRAM is
one-way during the execution, i.e., data only flows from the accelerator
toward DRAM. Therefore, the high latency of communicating with off-chip
DRAM does not stall the execution pipeline of the accelerator.
At the DRAM interface, messages are aggregated to form 512-bit lines, and are
written to the DRAM. A log index is kept at the DRAM interface
which is both used for writing new values to DRAM throughout the
execution, and reading valid values from the DRAM at the
end of execution.

\subsection{Runtime}
\label{SubSec:Runtime}

\begin{figure}
\begin{center}
\includegraphics[width=\linewidth,trim=0 12 0 0,clip]{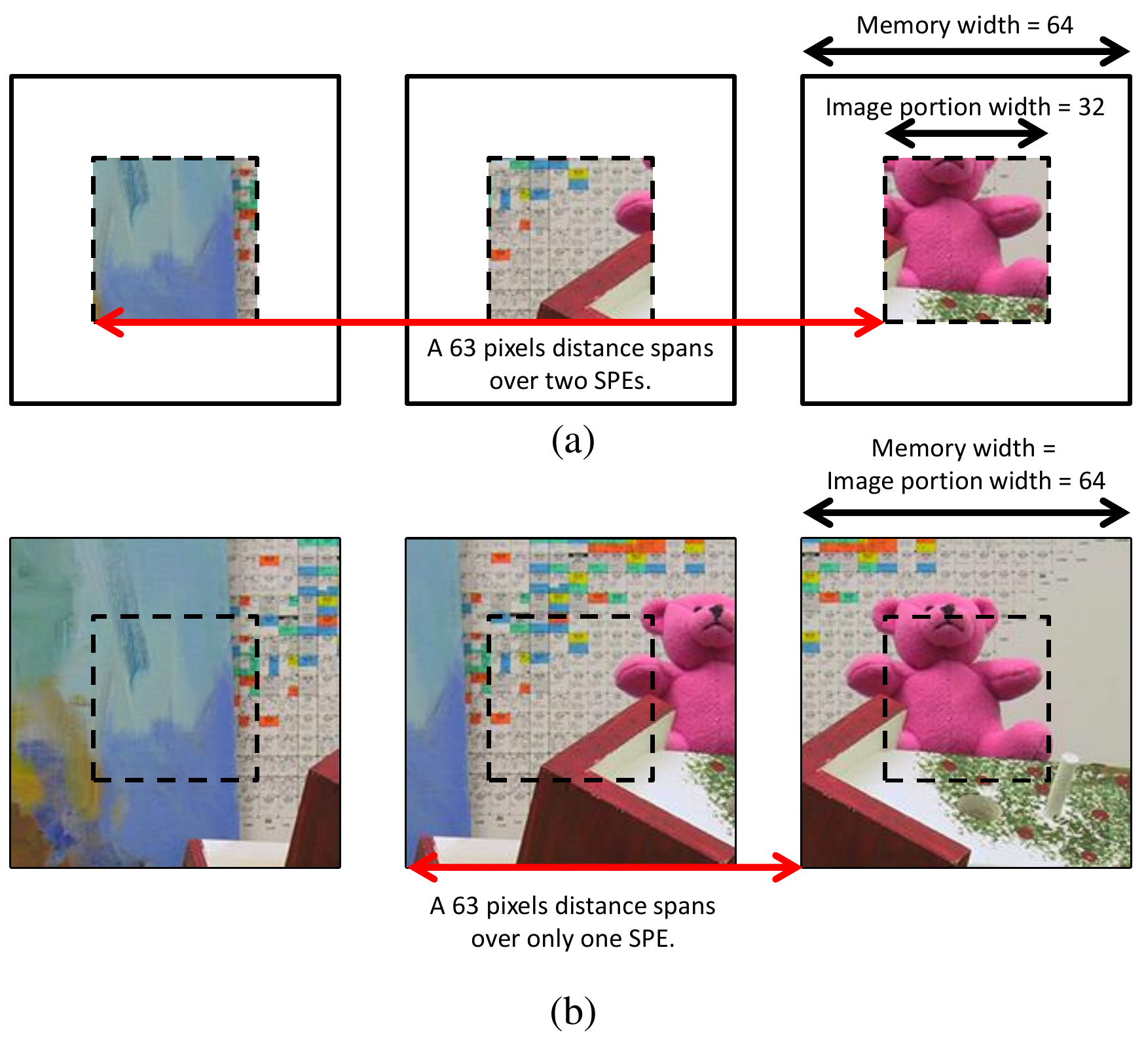}
\caption{(a) Example of an input set for stereo vision in which communications with SPEs farther than one hop is necessary to transfer singleton 2 data, and (b) solving this problem by replicating singleton 2 data.}
\label{Fig:SGL2Replication}
\end{center}
\end{figure}

The runtime is responsible for handling memory allocation,
parameter initialization, data padding when necessary, and
data placement and movement. 
Data padding might be
necessary depending on the input size, because work must be
distributed among SPEs evenly as the correct communication
of data between SPEs relies on this assumption.
Another assumption that our proposed communication scheme
builds upon is that all the data an SPE might possibly need,
whether label or singleton 2 data, must be available in at
most a single hop distance. Although this assumption always
holds for label data (only labels of
immediate neighboring RVs are needed), it might not necessarily
hold for singleton 2 depending on its access pattern and how
small the input data set is. 
Figure \ref{Fig:SGL2Replication}a illustrates this with an
example. In this example, the application is stereo vision \cite{StereoVision}
in which the singleton 2 accesses could reach 63 locations
to the left of any given RV. In this case, 
If the width of
the portion of the input assigned to each SPE is smaller than
the reach of singleton 2, then communication longer than
one hop will be necessary. Fortunately, replicating the
singleton 2 solves this problem,
as shown in Figure \ref{Fig:SGL2Replication}b for this example, 
and the runtime can implement this replication.

\subsection{Limitations and Future Work}
\label{SubSec:Limitations}

Some limitations of our proposed accelerator is inherent
to the specific Gibbs sampling algorithm selected, e.g., the lack
of support for continuous RVs. Some other limitations are
due to our design and implementation. For example, because
of the design choice to represent labels with six bits,
the proposed accelerator cannot support problem instances
with more than 64 labels. However, 64 labels is enough
for many applications \cite{StereoVisionTaxonomy, FPGAGibbs}, and expanding
the number of supported labels is future work. In addition,
previous work shows that slightly increasing bit width in
some places in the SPU datapath increases the result quality
to be closer to floating-point software implementations \cite{BeyondEndpoint}.
Incorporating those changes in our design is ongoing work,
but we expect the effects on area to be small.

Another limitation specific to our design is that it only supports
first-order MRF. Although this model can represent a wide
range of applications \cite{MRFVision, ComparativeStudyMRF, Stereophonic}, a more flexible label memory
design is required to expand the coverage to more applications
which we intend to do in the future.

Additionally, we plan to optimize the execution time
of MCMC by avoiding unnecessary RV updates. To be more specific, we can
skip a RV whose PDF is concentrated on only one value, i.e., there
is only one label to choose, and the labels of its neighborhood
has not changed. In other words, if a RV has a concentrated PDF,
its PDF will remain concentrated until something in its
neighborhood (i.e., the only changing input for MCMC update)
changes. Our preliminary experiments show that there is a reasonable
opportunity for improvement, and thus, adopting this optimization
is ongoing work.

%% file: 4-methodology.tex
\section{Methodology}
\label{Sec:Methodology}

\subsection{Applications and Metrics}
\label{SubSec:ApplicationsAndMetrics}

\begin{table}
  \small
  \centering
  \caption{Application parameters used in evaluations.}
  \label{Table:AcceleratorApplicationParameters}
  \begin{tabular}{|l|c|c|c|c|c|c|}
    \hline
    \multicolumn{7}{|c|}{\textbf{Motion Estimation}}\\
    \hline
    \hline
    & $\alpha$ & $\beta$ & $T$ & Labels & Size & Iters \\
    \hline
    dimetrodon &  &  &  &  & 584$\times$388 &  \\
    \cline{1-1}\cline{6-6}
    rubberwhale & 6 & 6 & 1 & 49 & 584$\times$388 & 3000 \\
    \cline{1-1}\cline{6-6}
    venus &  &  &  &  & 210$\times$190 &  \\
    \hline
    \hline
    \multicolumn{7}{|c|}{\textbf{Stereo Vision}}\\
    \hline
    \hline
    & $\alpha$ & $\beta$ & $T$ & Labels & Size & Iters \\
    \hline
    art &  &  &  & 28 & 348$\times$278 & 3000 \\
    \cline{1-1}\cline{5-7}
    poster & 6 & 7 & 2 & 30 & 435$\times$383 & 1500 \\
    \cline{1-1}\cline{5-7}
    teddy &  &  &  & 56 & 450$\times$375 & 3000 \\
    \hline
  \end{tabular}
\end{table}

We use two image analysis applications, namely, motion
estimation \cite{MotionEstimation} and
stereo vision \cite{StereoVision, StereoVisionTaxonomy}
to evaluate our design in the optimization mode. Motion
estimation is covered in detail in Section \ref{SubSec:MotionEstimation}.
Stereo vision reconstructs the depth information of objects
in a field captured from two cameras by matching the pixels
between the two images. The farther the location of the pixel
in the two images, the deeper it is in the field. Therefore,
singleton 1 data comes from the right view, and singleton 2
comes from each of the $L$ pixels preceding the target pixel
in the left view, where $L$ is the number of labels in the model.

We evaluate each application using three input image sets from
Middlebury \cite{MiddleburyMotionEstimation, StereoVisionTaxonomy}.
Table \ref{Table:AcceleratorApplicationParameters} summarizes the parameters used for each input.
Parameter names correspond to those in Figure \ref{Fig:MCMC}.
To generate outputs, we calculate the mode of the labels
in the last 1,000 iterations for each input.
We compare the results against a MATLAB implementation
which uses double-precision floating-point to assess the
quality of the results. We use end-point
error (EPE) as the metric for evaluating motion estimation results \cite{MotionEstimation},
and bad-pixel (BP) percentage as the metric for stereo vision \cite{StereoVisionTaxonomy}.

\begin{figure}
\begin{center}
\includegraphics[width=\linewidth,trim=10 160 10 100,clip]{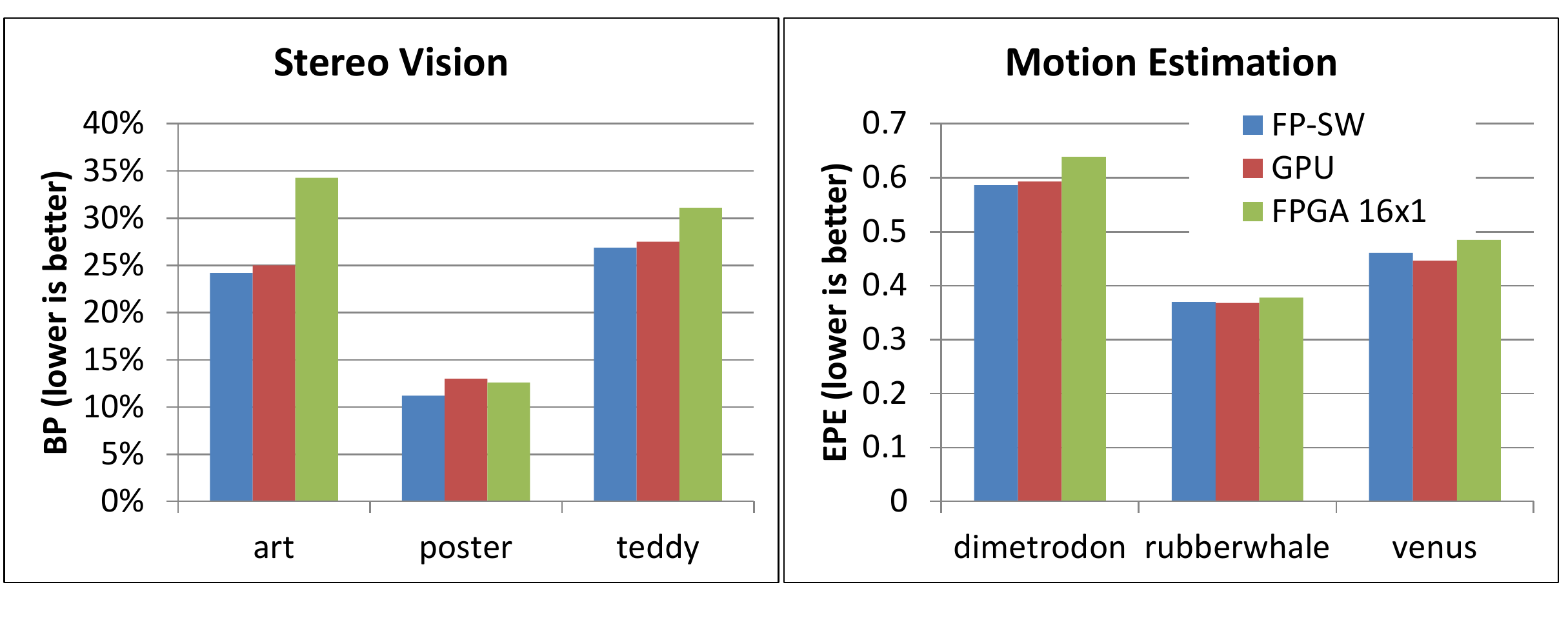}
\caption{Comparison of output quality results between a MATLAB floating-point implementation, a GPU implementation, and our FPGA prototype.}
\label{Fig:QualityResults}
\end{center}
\end{figure}

\subsection{HLS Implementation}
\label{SubSec:HLSImplementation}
To implement the FPGA prototype and perform ASIC analysis of
our proposed accelerator, we use High-Level Synthesis (HLS) to
compile code written in C++ to Hardware Description Language (HDL).
We utilize Intel HLS compiler from Quartus 18.0 \cite{HLS} to
implement the FPGA prototype. We implement the components
shown in Figure \ref{Fig:SPE} individually, and then connect
them together using Platform Designer \cite{QSys}, and synthesize
the final design for a Programmable Acceleration Card (PAC)
with Arria 10 GX FPGA \cite{PACArria10} using Quartus 17.1.
We utilize Open Programmable Acceleration Engine (OPAE) 1.2
to develop the runtime that controls the
FPGA prototype.

For ASIC analysis, we use Mentor Catapult \cite{Catapult}
and adapt our C++ code to use Algorithmic C datatypes \cite{ACDataTypes} which
allow for using custom precision data types in the HLS design.
We utilize Design Compiler \cite{DesignCompiler} to synthesize our design using
a 15nm library \cite{Nangate-15} to derive area and power results
for non-memory logic. In addition, we use CACTI 7.0 \cite{CACTI} to
estimate the area and power of memory components. Since the smallest
technology node in CACTI is 22nm, we conservatively use those
numbers for area and power calculation. Power numbers
are calculated by feeding the switching activity based on a 32-label
application to Design Compiler, conservatively assuming all
input ports switch every time new data arrives.

\subsection{GPU Implementation}
\label{SubSec:GPUImplementation}

We implement the two applications using
CUDA \cite{CUDASDK}, and conduct evaluations on an Nvidia
RTX 2080 Ti GPU \cite{RTX2080}. The same chromatic schedule for updating
conditionally independent RVs in parallel is used in
the GPU implementation. We applied spatial-tiling \cite{Tiling} to take
advantage of spatial locality, i.e., we divided the input image into
equal-sized rectangles and assigned each region to a specific thread
block. The size of the thread blocks were 16$\times$16, which means they
covered a 32$\times$16 region (due to the chromatic schedule we use for updates).

%% file: 5-evaluation.tex
\section{Evaluation}
\label{Sec:Evaluation}

\subsection{FPGA Prototype}
\label{SubSec:FPGAPrototype}

\subsubsection{Result Quality} ~
\label{SubSubSec:PrototypeResultQuality}

Figure \ref{Fig:QualityResults} demonstrates the application result quality for the two applications discussed in Section \ref{SubSec:ApplicationsAndMetrics},
using their corresponding metrics. The results are consistent with
prior work \cite{SPU}.
It is possible to further improve the
quality of the results by slightly modifying the bit width
of some places in the SPU datapath, which is discussed
in detail elsewhere \cite{SPU}.

\subsubsection{Performance}
\label{SubSubSec:PrototypePerformance}

Resource requirements and clock rate for two design points
on an Intel Arria 10 GX FPGA are presented in Table \ref{Table:FPGAResults}.
(16$\times$2 means 16 SPEs and two SPUs/SPE.)
These FPGA implementations support up to 256K RVs, big enough
for the input datasets used in our evaluations.
We only implement designs with one and two SPU(s)/SPE,
because as discussed in Section \ref{SubSubSec:LabelMemory},
accesses to LMem are pipelined for SPUs in the same SPE.
The implication is that if an application has less labels than there are
SPUs in an SPE, then additional SPUs will not be utilized. Since
we can guarantee that all applications have at least two labels
(otherwise there would be no problem to solve), we implement designs
with at most two SPUs/SPE.

\begin{figure}
\begin{center}
\includegraphics[width=\linewidth,trim=10 160 10 115,clip]{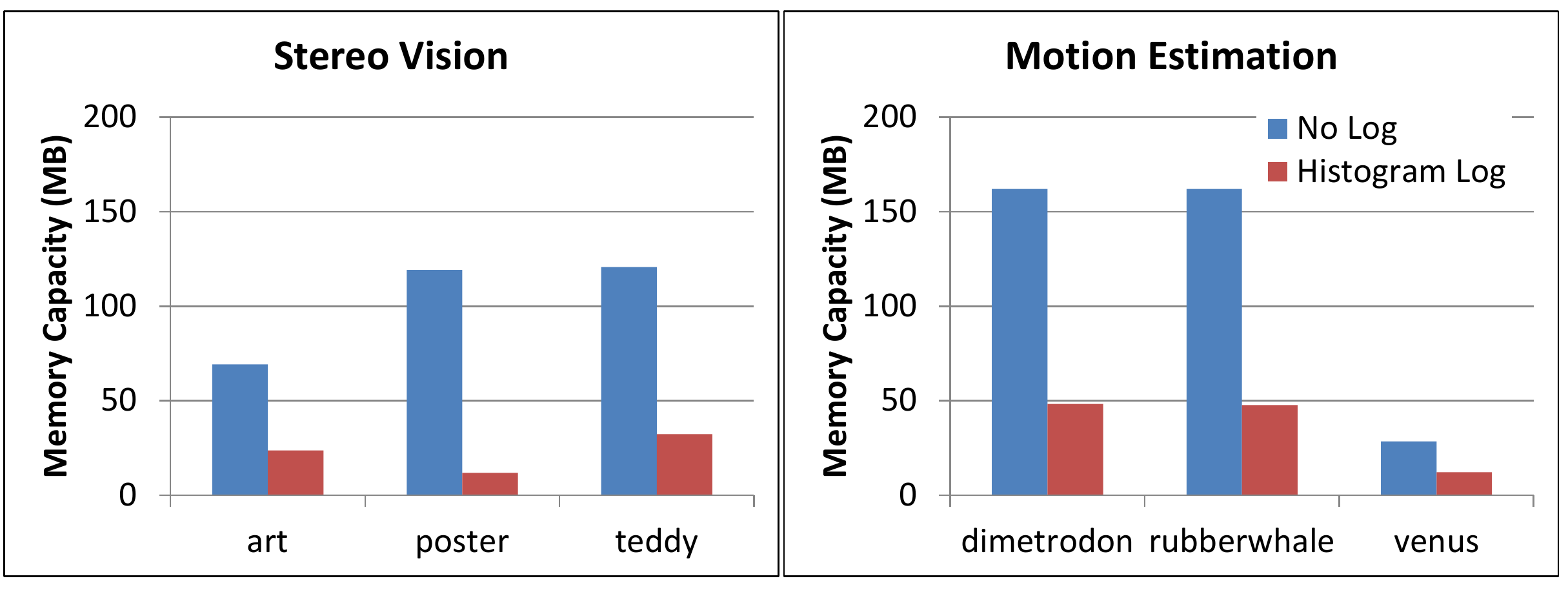}
\caption{Memory space to store data for generating labels histogram. ``No Log'': results written to off-chip memory every iteration. ``Histogram Log'': off-chip capacity to store the log, and the on-chip memory storing the labels and their corresponding counters.}
\label{Fig:HistogramStorage}
\end{center}
\end{figure}

\begin{table}
  \small
  \centering
  \begin{threeparttable}
  \caption{FPGA prototypes results.}
  \label{Table:FPGAResults}
  \begin{tabular} {|c|c|c|c| }
    \hline
    Design Point & 16$\times$1 & 16$\times$2 & \textbf{Device Max}  \\
    \hline
    \hline
    ALM\tnote{1} & 182,690 & 247,815 & \textbf{427,200} \\
    \hline
    M20K\tnote{2} & 1,376 & 1,545 & \textbf{2,713} \\
    \hline
    DSP & 160 & 320 & \textbf{1,518} \\
    \hline
    Clock Rate & 185MHz & 146MHz & \textbf{667MHz} \\
    \hline
    Total Perf.\tnote{3} & 2.96B & 4.672B & \\
    \hline
  \end{tabular}
  \begin{tablenotes}
      \item [$^{\tiny 1}$Adaptive logic module. $^{\tiny 2}$ 20K-bit memory block. $^{\tiny 3}$ Labels/sec.]
  \end{tablenotes}
  \end{threeparttable}
\end{table}

As it is expected, the design with two SPUs/SPE occupies less than
twice the area of the other design, which means the area of some
components (e.g., scheduler, memory control, etc.) are successfully
amortized. However, the clock rate drops due to the more complicated
routing required between the SPEs.

Equation \ref{Eq:FPGASpeedup} shows that compared with prior work
on FPGAs \cite{FPGAGibbs}, our proposed accelerator (16$\times$2 design point)
achieves 26$\times$ speedup. (See Table II in \cite{FPGAGibbs}.)
This is mainly due to
the better memory design and avoiding off-chip communication
as much as possible. Nevertheless, superior performance is not the
only advantage of our work. Due to our proposed tiled architecture, efficient
memory system design, and incorporating the scheduling logic into SPEs,
our design provides far more flexibility compared
to \cite{FPGAGibbs}, which only supports MRF models up to a
certain row size.
\begin{equation}
    \frac{4.672 * 10^9 labels/sec}{2 labels/sample * 88.588 * 10^6 samples/sec} = 26.37
    \label{Eq:FPGASpeedup}
\end{equation}

\subsubsection{Uncertainty Quantification}
\label{SubSubSec:PrototypeUQ}

The amount of memory used to store information for generating the labels
histogram in our FPGA prototype and a hypothetical design which
stores all labels in off-chip memory are compared
in Figure \ref{Fig:HistogramStorage}.
Another baseline would be a design that has
a counter for each possible label of each RV. However, if the
counters reside on the chip, they require enormous area
(e.g., with 10-bit counters, 80 bytes for each RVs) which will not
be utilized for applications with less than 64 labels. Even for
applications with a large number of labels, our analysis (Figure \ref{Fig:UniqueLabelsMotionEstimation})
shows that only a few unique labels are chosen throughout the
execution. Moreover, if the counters are stored in off-chip memory,
two-way communication is needed to update the counts. Therefore, we do not include it
in our comparisons. The figure shows that our hybrid on-chip/off-chip
memory system and logging scheme saves an average of 71\% in memory space
for generating the histogram.

\subsection{ASIC Analysis}
\label{SubSec:ASIC}

\begin{figure}
\begin{center}
\includegraphics[width=\linewidth,trim=50 150 50 150,clip]{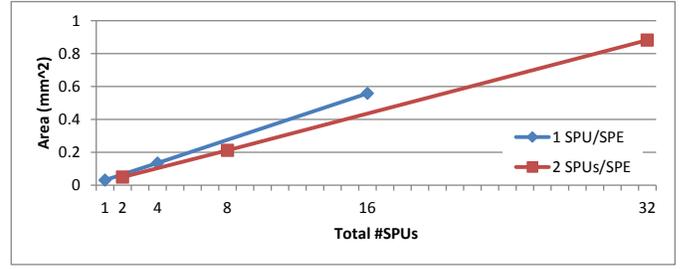}
\caption{Accelerator area at different design points.}
\label{Fig:ASICAreaScaling}
\end{center}
\end{figure}

\begin{figure}
\begin{center}
\includegraphics[width=.85\linewidth,trim=0 130 0 50,clip]{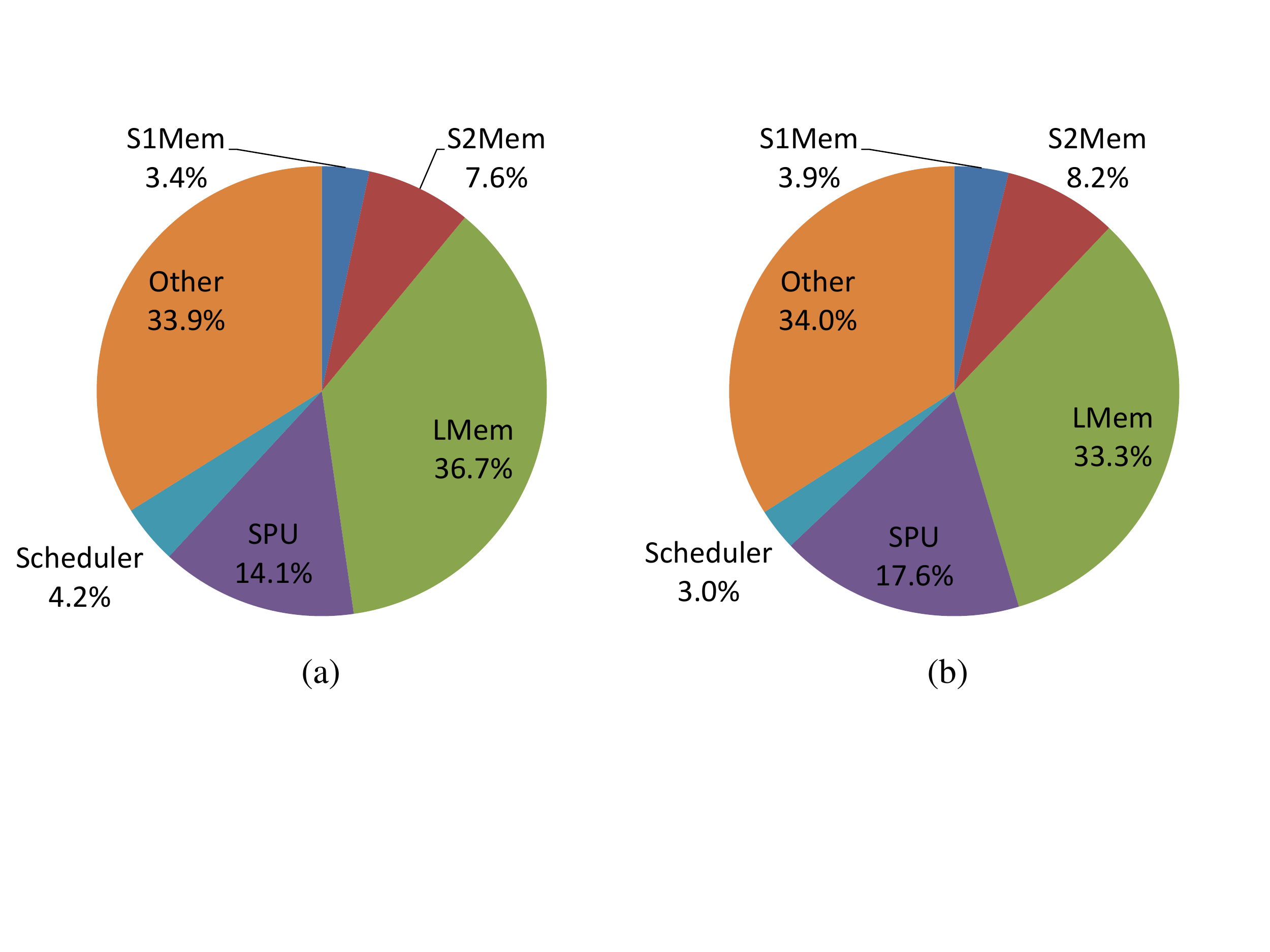}
\caption{SPE area breakdown with (a) one SPU, and (b) two SPUs. Total area: (a) 30,760$\mu m^2$, and (b) 48,862$\mu m^2$.}
\label{Fig:ASICComponentArea}
\end{center}
\end{figure}

\begin{figure}
\begin{center}
\includegraphics[width=.85\linewidth,trim=0 130 0 50,clip]{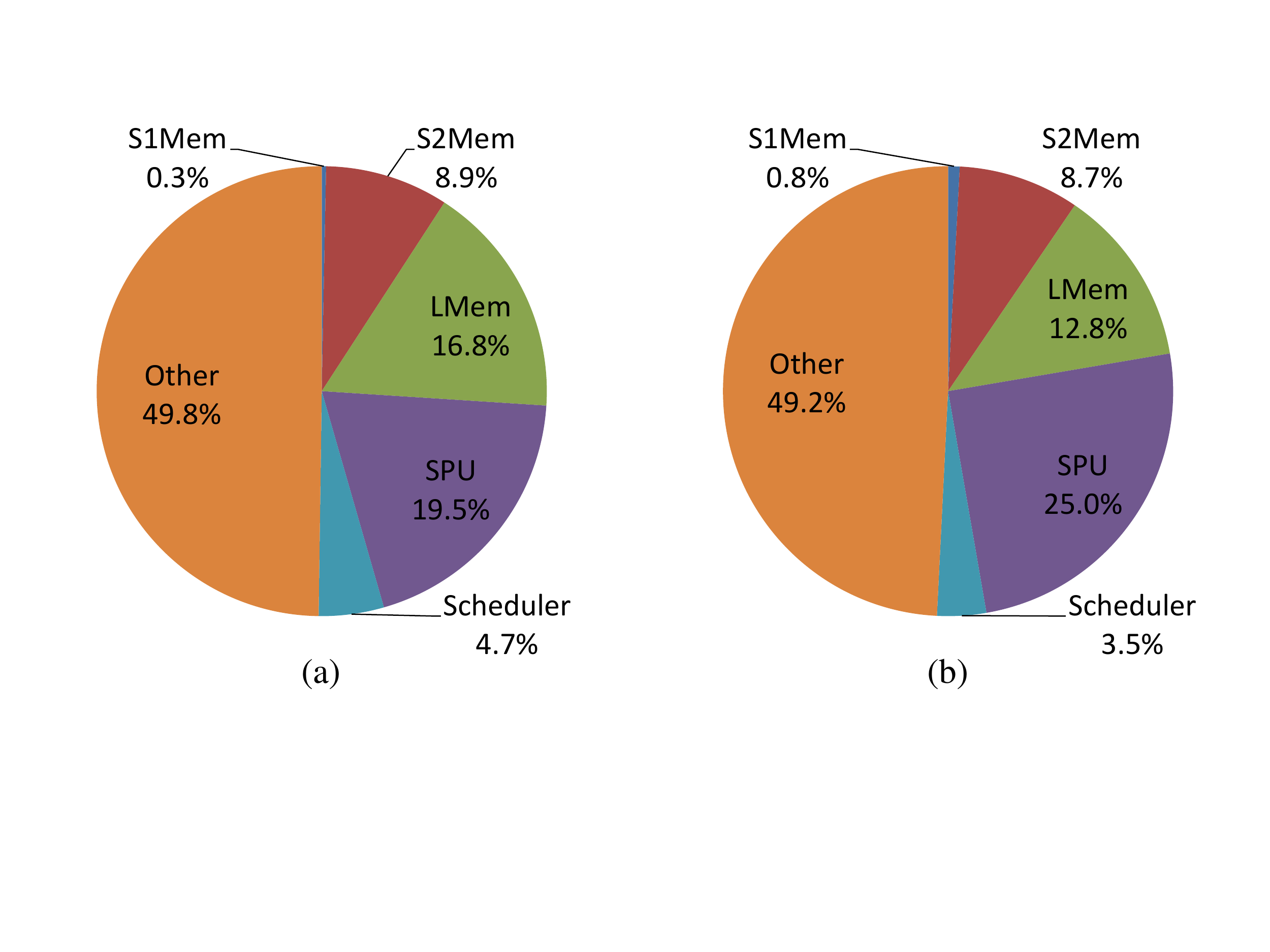}
\caption{SPE power breakdown with (a) one SPU, and (b) two SPUs. Total power: (a) 50.076mW, and (b) 78.917mW.}
\label{Fig:ASICComponentPower}
\end{center}
\end{figure}

We implement and synthesize ASIC designs with one, four,
and 16 SPEs, each with one and two SPU(s)/SPE. Figures \ref{Fig:ASICComponentArea}-\ref{Fig:ASICComponentPower}
show the area and power breakdown by component of an SPE with one and two SPU(s).
Compared with the one SPU/SPE design,
we observe the area and power amortization trend in two SPUs/SPE design for ASIC
designs too. The design with two SPUs/SPE uses 20.6\% less
area and 21.2\% less power per SPU compared to that with one SPU/SPE.
In addition, another indicator of successful overhead amortization is that
the fraction of area and power used by SPUs, which perform
the main computations, increases with two SPUs/SPE by 24.8\% and 28.2\%, respectively.
(Note that these are post-synthesis results,
place and route might yield different numbers).

The overall accelerator
area results for multiple design points are depicted in Figure \ref{Fig:ASICAreaScaling}.
In each design, 1K RV is assigned
to each SPU, i.e., a design with 2048 SPEs with 1SPU/SPE
supports the same amount of memory as a design with 1024 SPEs
with 2 SPUs/SPE, and both support Full-HD images.
As expected, due to the homogeneity of the proposed tiled architecture,
the area scales almost linearly with the number of SPEs.
We predict the area of an
accelerator with 1024 SPEs, each with two SPUs by extrapolating
this graph and adding the area of the required DRAM hubs
to be 58$mm^2$, i.e., 92.3\% smaller than an RTX 2080 Ti GPU (754$mm^2$) \cite{RTX2080}.

\begin{table*}
  \small
  \centering
  \begin{threeparttable}
  \caption{Qualitative comparison of our proposed accelerator to other Bayesian inference accelerators.}
  \label{Table:QualitativeComparison}
  \begin{tabular}{|l|c|c|c|c|}
    \hline
    \textbf{} & \textbf{Application Flexibility} & \textbf{Input Flexibility} & \textbf{Memory System} & \textbf{Uncertainty Quantification} \\
    \hline
    \hline
    Gibbs tile \cite{EricJonas} & Medium & High & On-chip & \ding{53} \\
    \hline
    SPU \cite{SPU} & Medium & -- & -- & -- \\
    \hline
    FlexGibbs \cite{FPGAGibbs, FlexGibbs} & Medium & Low & Off-chip & High Overhead \\
    \hline
    VIP \cite{VIP} & High & Very High & 3D-stacked & \ding{53} \\
    \hline
    AcMC$^2$ \cite{AcMC2} & Very High & Very High & Off-chip & Trades Accuracy for Memory Capacity \\
    \hline
    \hline
    \textbf{This work} & \textbf{Medium} & \textbf{High} & \textbf{Hybrid On-chip/Off-chip} & \textbf{Efficient and Accurate} \\
    \hline
  \end{tabular}
  \end{threeparttable}
\end{table*}

\begin{figure}
\begin{center}
\includegraphics[width=\linewidth,trim=50 150 50 150,clip]{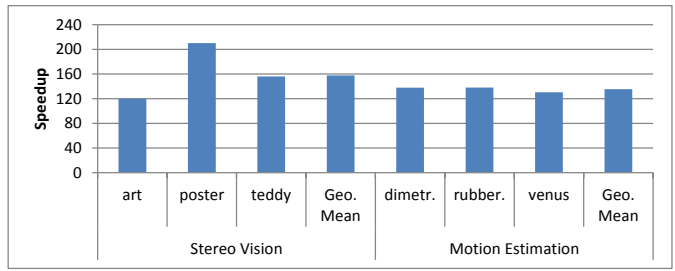}
\caption{Speedup of a 1024$\times$2 SPE ASIC design at 15nm, compared to an RTX 2080 Ti GPU for stereo vision and motion estimation. GPU execution times in ms: \texttt{art}: 242, \texttt{poster}: 303, \texttt{teddy}: 839, \texttt{dimetrodon}: 1082, \texttt{rubberwhale}: 1082, \texttt{venus}: 204.}
\label{Fig:GPUASICComparison}
\end{center}
\end{figure}

In terms of performance, compared to an RTX 2080 Ti,
the aforementioned accelerator achieves speedups in the range of
120$\times$-210$\times$, with geometric means of
135$\times$ and 158$\times$ for motion estimation and
stereo vision, respectively, as demonstrated in Figure \ref{Fig:GPUASICComparison}.
The implication of these numbers is that
our proposed accelerator can process Full-HD images at 30fps
with 64 labels for 1500 iterations per frame.

%% file: 6-relatedwork.tex
\section{Related Work}
\label{Sec:RelatedWork}

Methods for parallelizing Gibbs sampling based on the
conditional independence of RVs
have been proposed \cite{EricJonas, ParallelGibbs, GPUAcceleratedGibbs}. Our work takes advantage
of similar principles to create a schedule to
update conditionally independent RVs in parallel.

So et al. \cite{CustomLayout} present a custom data layout approach in multiple
memory banks for array-based computations.
Wang et al. \cite{Polyhedral} propose a polyhedral model that attempts to
detect memory bank conflicts for generalized memory
partitioning in HLS. Cilardo and Gallo \cite{LatticeBased} present a lattice-based method
that takes advantage of the Z-polyhedral model \cite{ZPolyhedral} for
program analysis and adopt a partitioning scheme
based on integer lattices. Escobedo and Lin \cite{Tessellating}
use memory space tessellation to find patterns in data
accesses and cover the
memory access space with the found pattern.
In other works, Escobedo and Lin \cite{NonStencilGraphColoring, StencilGraphColoring}
present approaches that create the data dependence graph of memory
accesses in the iteration domain, and use graph coloring to assign
data elements to memory banks.
These works address the problem of determining the proper memory structure for
a specific problem that uses HLS, whereas our goal in this
paper, while being an instance of this problem, is to design
a fixed memory structure that can support a wide range of
applications. Moreover, in these works, the communication
among different compute units is not accounted for, which
could further constrain the data placement solution.

Table \ref{Table:QualitativeComparison} presents a summary of
the differences between our accelerator and some of the related work.
\footnote{Application and Input Flexibility in the table refer to the diversity
of applications and input sizes supported.}
To the best of our knowledge, other MCMC and inference accelerators
in the literature do not address the memory subsystem
challenges and uncertainty quantification as
comprehensively and effectively as this work. 
Jonas \cite{EricJonas} presents a tiled Gibbs sampling architecture,
but does not include an efficient memory system
design and support for uncertainty quantification. Zhang et al. \cite{SPU} propose
and analyze a microarchitecture for a Gibbs sampling
function unit, but does not propose an actual accelerator design.
and only includes back of the envelope calculations
for performance. 
Ko et al. \cite{FlexGibbs, FPGAGibbs} design
an FPGA-based parallel Gibbs sampling accelerator for
MRF, which does not provide
the level of flexibility supported in this work. Ko et al. \cite{PGMA} propose
a similar architecture for a 2-core ASIC accelerator, with
the addition of asynchronous Gibbs sampling (AGS). This change
reduces the accuracy of the algorithm, which requires more
iterations to achieve similar result quality, but also results
in lower memory capacity requirement on the chip.
Hurkat and Mart\'inez \cite{VIP} propose a vector
processor for deterministic inference algorithms
which utilizes 3-D stacking to address
high memory bandwidth requirements. Banerjee et al. \cite{AcMC2} design a
compiler that transforms probabilistic models into hardware
accelerators. Although their
work supports more general models, it produces a new
accelerator per model and is different
from our work, in that our goal is to design
an accelerator that supports a reasonable range of problems.
Additionally, support for uncertainty quantification
in their work is limited due to the use of on-chip counters, which can
impose significant overheads when the problem size grows,
and the adoption of binning and Bloom filters \cite{BloomFilter} to approximate
the histogram when the number of entries is high. Mingas and Bouganis present a
streaming FPGA architecture to accelerate Parallel Tempering MCMC, an MCMC
method to sample from multi-modal distributions, that uses custom precision
without introducing sampling errors \cite{ParallelTempering}. Their work, however,
is not a parallel architecture and is more suitable for addressing multi-modality.
Rouhani et al. \cite{CausaLearn} present an FPGA accelerator for Hamiltonian Monte-Carlo
which relies on initial profiling to customize hardware resource allocation and scheduling
for complex streaming scenarios. Their work solves problems with continous RV spaces,
but it lacks uncertainty quantification.

%% file: 7-conclusion.tex
\section{Conclusion}
\label{Sec:Conclusion}

Probabilistic algorithms, such as MCMC, are an attractive approach in
statistical machine learning which offer interpretability
and uncertainty quantification of the final results. These algorithms, however,
require probabilistic computations which are not a good fit for conventional
processors. We propose a specialized accelerator to significantly improve
the performance of MRF inference using MCMC compared to general-purpose
processors. 
Our proposed architecture takes advantage of near-memory
computing, memory banking, and communication schemes tailored
to the characteristics of first-order MRF model. Importantly, we introduce novel memory
system support for uncertainty quantification by employing a hybrid on-chip/off-chip
memory system. We prototype the proposed design with 32 function units
on an Arria 10 FPGA using Intel HLS compiler and achieve a 146MHz clock
rate. The FPGA implementation outperforms the previous work by 26$\times$.
ASIC analysis using Mentor Graphics HLS compiler shows that in 15nm technology,
the accelerator runs at 3GHz and achieves 120$\times$-210$\times$
speedup over GPU implementations of motion estimation and stereo vision.

This work takes an important first step by demonstrating an
accelerator for probabilistic algorithms with efficient uncertainty quantification. Further exploration of the trade-offs between
specialization and generalization is an interesting endeavor and we will
continue our work in this space in the future.